\documentclass{pnastwo}

\usepackage[pdftex]{graphicx}

\usepackage[usenames,dvipsnames]{color}
\usepackage{float}
\definecolor{ao(english)}{rgb}{0.0, 0.5, 0.0}
\definecolor{forestgreen(web)}{rgb}{0.13, 0.55, 0.13}

\usepackage[normalem]{ulem}

\usepackage{amssymb,amsfonts,amsmath}
\usepackage{calc}

\begin{document}


\title{Flagellated bacterial motility in polymer solutions}



\author{V. A. Martinez\affil{1}{SUPA, School of Physics and Astronomy, The University of Edinburgh, King's Buildings, Mayfield Road, Edinburgh EH9 3JZ, UK}, J. Schwarz-Linek\affil{1}{}, M. Reufer\affil{1}{}, L. G. Wilson\affil{2}{Department of Physics, University of York, York YO10 5DD, UK}, A. N. Morozov\affil{1}{}, W. C. K. Poon\affil{1}{}}

\contributor{Submitted to Proceedings of the National Academy of Sciences of the United States of America}

\maketitle

\begin{article}

\begin{abstract} 
It is widely believed that the swimming speed, $v$, of many flagellated bacteria is a non-monotonic function of the concentration, $c$, of high-molecular-weight linear polymers in aqueous solution, showing peaked $v(c)$ curves. Pores in the polymer solution were suggested as the explanation. Quantifying this picture led to a theory that predicted peaked $v(c)$ curves. Using new, high-throughput methods for characterising motility, we have measured $v$, and the angular frequency of cell-body rotation, $\Omega$, of motile \textit{Escherichia coli} as a function of polymer concentration in polyvinylpyrrolidone (PVP) and Ficoll solutions of different molecular weights. We find that non-monotonic $v(c)$ curves are typically due to low-molecular weight impurities. After purification by dialysis, the measured $v(c)$ and $\Omega(c)$ relations for all but the highest molecular weight PVP can be described in detail by Newtonian hydrodynamics. There is clear evidence for non-Newtonian effects in the highest molecular weight PVP solution. Calculations  suggest that this is due to the fast-rotating flagella `seeing' a lower viscosity than the cell body, so that flagella can be seen as nano-rheometers for probing the non-Newtonian behavior of high polymer solutions on a molecular scale. 
\end{abstract}

\keywords{Bacterial motility | Escherichia coli | Non-Newtonian | Polymer solution}



\dropcap{T}he motility of micro-organisms in polymer solutions is a topic of vital biomedical interest. Thus, e.g., mucus covers the respiratory \cite{RTmucin},  gastrointestinal \cite{GImucin} and reproductive \cite{spermMucus} tracks of all metazoans. Penetration of this solution of biomacromolecules by motile bacterial pathogens is implicated in a range of diseases, e.g., stomach ulcers  caused by {\it Helicobacter pylori} \cite{pylori}. Oviduct mucus in hens provides a barrier against {\it Salmonella} infection of eggs  \cite{JAVMA}. Penetration of the exopolysaccharide matrix of biofilms by swimming bacteria \cite{Briandet2012} can stabilise or destabilise them {\it in vivo} (e.g. the bladder) and {\it in vitro} (e.g. catheters). In reproductive medicine (human and veterinary), the motion of sperms in seminal plasma and vaginal mucus, both non-Newtonian polymer solutions, is a strong determinant of fertility \cite{spermMucus}, and polymeric media are often used to deliver spermicidal and other vaginal drugs \cite{AAPS}.

Micro-organismic propulsion in non-Newtonian media such as high-polymer solutions is also a `hot topic' in biophysics, soft matter physics, and fluid dynamics \cite{LaugaPowers}. Building on  knowledge of propulsion  modes at low Reynolds number in Newtonian fluids \cite{LaugaPowers}, current work seeks to understand how these are modified to enable efficient non-Newtonian swimming. In particular, there is significant interest in  a flapping sheet \cite{LaugaSheet,ShelleySheet} or an undulating filament \cite{FuFilament} (modelling the sperm tail)  and in a rotating rigid helix (modelling the flagella of, e.g., {\it Escherichia coli}) \cite{LeshanskyHelix,BreuerHelix} in non-Newtonian fluids. 

An influential set of experiments in this field was performed 40 years ago by Schneider and Doetsch (SD) \cite{Schneider74}, who measured the average speed, $\bar v$, of seven flagellated bacterial species  (including {\it E. coli}) in solutions of polyvinylpyrrolidone (PVP, molecular weight given as $M = 360$kD) and in methyl cellulose (MC, $M$ unspecified) at various concentrations, $c$. SD claimed that  $\bar v(c)$ was always non-monotonic and peaked.

A qualitative explanation was suggested by Berg and Turner (BT) \cite{Berg79}, who argued that entangled linear polymers formed `a loose quasi-rigid network easily penetrated by particles of microscopic size'. BT measured the angular speed, $\Omega$, of the rotating bodies of tethered {\it E. coli} cells in MC solutions. They found that adding MC hardly decreased $\Omega$. However, in solutions of Ficoll, a branched polymer, $\Omega\propto \eta^{-1}$, where $\eta$ is the solution's viscosity, which was taken as evidence for Newtonian behavior. In MC solutions, however, BT suggested that there were {\it E. coli}-sized pores, so that cells rotated locally in nearly pure solvent. Magariyama and Kudo (MK) \cite{Magariyama02} formulated a theory based on this picture, and predicted a peak in $v(c)$ by assuming that a slender body in a linear-polymer solution experienced different viscosities for tangential and normal motions in BT's `easily penetrated' pores. 

This `standard model' is widely accepted in the biomedical literature on flagellated bacteria in polymeric media. It also motivates much current physics research in non-Newtonian low-Reynolds-number propulsion. Nevertheless, there are several reasons for a fundamental re-examination of the topic. 

First, polymer physics \cite{Colby} casts {\it a priori} doubt on the presence of {\it E. coli}-sized pores in an entangled solution.  Entanglement occurs above the `overlap concentration', $c^*$, where coils begin to touch. The `mesh size' at $c^*$, comparable to a coil's radius of gyration, $r_g$, gives the maximum possible pore size in the entangled network. For 360kD PVP in water, $r_g \lesssim 60$~nm (see below), well under the cross section of  {\it E. coli}  (0.8$\,\mu$m). Thus, the physical picture suggested by BT \cite{Berg79} and used by MK \cite{Magariyama02} has doubtful validity. \\

\setlength{\parindent}{0cm}
\colorbox{SkyBlue}{
\fbox{
	\parbox{8cm}{
	
	{\color{blue}{\bf Significance} 

The way micro-organisms swim in concentrated polymer solutions has important biomedical implications, e.g. that is how pathogens invade the mucosal lining of mammal guts. Physicists are also fascinated by self-propulsion in such complex, `non-Newtonian' fluids. The current `standard model' of how bacteria propelled by rotary helical flagella swim through concentrated polymer solutions postulates bacteria-sized `pores', allowing them relative easy passage. Our experiments using novel high-throughput methods overturn this `standard model'. Instead, we show that the peculiarities of flagellated bacteria locomotion in concentrated polymer solutions are due to the fast-rotating flagellum, giving rise to a lower local viscosity in its vicinity. The bacterial flagellum is therefore a `nano-rheometer' for probing flows at the molecular level.}}
}
}\\ 

\setlength{\parindent}{.6cm}

Secondly, SD's data were statistically problematic. They took movies, from which cells with `the 10 greatest velocities were used to calculate the average velocity' \cite{Schneider74}. Thus, their `peaks' in $v(c)$ could be no more than fluctuations in measurements that were in any case systematically biased.

Finally, while MK's theory indeed predicts a peak in $v(c)$, we find that their formulae also predict a monotonic increase in $\Omega(c)$ in the same range of $c$ (Fig.~S1), which is inconsistent with the data of BT, who observed a monotonic decrease. 

We have therefore performed a fresh experimental study of {\it E.~coli} motility using the same polymer (PVP) as SD, but varying the molecular weight, $M$, systematically. High-throughput methods for determining $v$ and $\Omega$ enabled us to average over $\sim 10^4$ cells at each data point. Using polymers as purchased, we indeed found peaked $v(c)$ curves at all $M$ studied. However, purifying the polymers removed the peak in all but a single case. Newtonian hydrodynamics can account in detail for the majority of our results, collapsing data onto master curves. We show that the ratio $v(c)/\Omega(c)$ is a sensitive indicator of  non-Newtonian effects, which we uncover for 360kD PVP. We argue that these are due to shear-induced changes in the polymer around the flagella.

Below, we first give the necessary theoretical and experimental background before reporting our results. 

\section{Theoretical groundwork: solving Purcell's model}

\begin{figure}[h]
\vspace*{-.5cm}
	\begin{center}
	\includegraphics[width=2in]{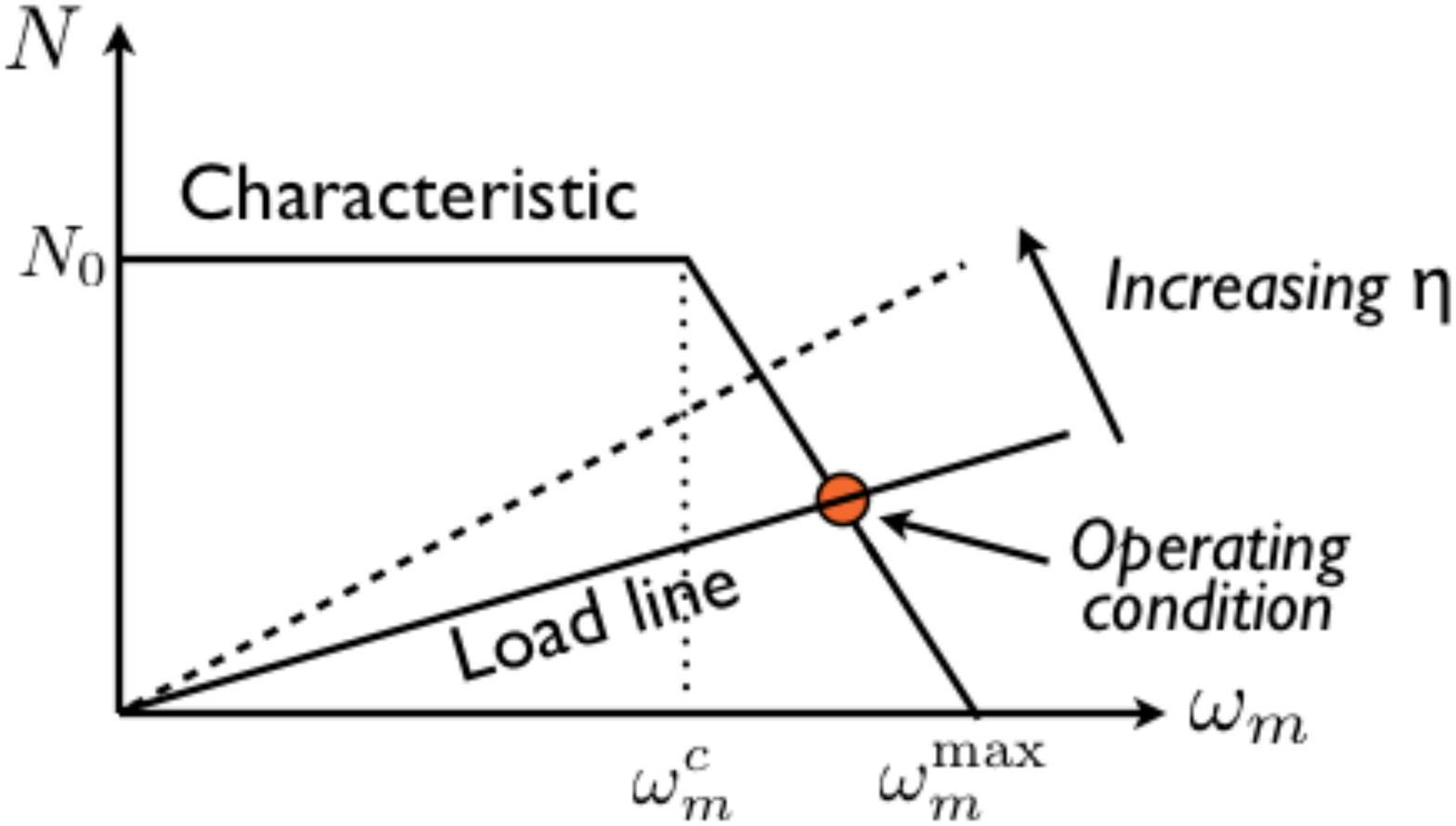}
	\caption{Schematic of the relationship $N(\omega_m)$ between the flagellum motor torque, $N$, and its angular speed, $\omega_m$. Intersection with a load-line determines the operating condition. The $N(\Omega)$ relation has the same form (cf. Eq.~\ref{motor}).}
	\label{loadline}
	\end{center}
\end{figure}

Purcell's widely-used `model {\it E. coli}' has a prolate ellipsoidal cell body bearing a single left-handed helical flagellum at one pole \cite{Purcell97}. Its motion is described by three kinematic parameters: the swimming speed, $v$,  the body angular speed, $\Omega$, and the flagellum angular speed, $\omega$:
\begin{equation}
\mathbf{v} = (v,0,0), \, \boldsymbol{\omega} = (-\omega,0,0), \, \mathbf{\Omega} = (\Omega,0,0) \label{components},
\end{equation}
with $v, \omega, \Omega > 0$. The drag forces and torques ($\mathbb{F}, \mathbb{N}$) on the body (subscript `$b$') and flagellum (subscript `$f'$) are given by
\begin{eqnarray}
\left( \begin{array}{cc}
\mathbb{F}_b \\
\mathbb{N}_b
\end{array}
\right) & = & 
-\left( \begin{array}{cc}
A_0 & 0 \\
0& D_0 \\
\end{array}
\right)
\left( \begin{array}{cc}
\mathbf{v} \\
\mathbf{\Omega}
\end{array}
\right)  \label{bodyFN}\\
\left( \begin{array}{cc}
\mathbb{F}_f \\
\mathbb{N}_f
\end{array}
\right) & = &
-\left( \begin{array}{cc}
A & B \\
B & D \\
\end{array}
\right)
\left( \begin{array}{cc}
\mathbf{v} \\
\boldsymbol{\omega}
\end{array}
\right), \label{helixFN}
\end{eqnarray}
where $A_0, D_0, A, B, D \propto \eta_s$, the solvent viscosity. Requiring the body and flagellum to be force and torque free, we find
\begin{eqnarray}
\Omega &=& {\frac { D\left(A_{{0}}+
A \right)-{B}^{2}}{D_{{0}} \left( A + A_0 \right) }}\omega \equiv \beta \omega, \label{Omega_omega}\\
 v &=& {\frac {B}{A_{{0}}+A}} \omega\equiv \gamma \omega, \label{v_omega}
\end{eqnarray}
where $\beta$ and $\gamma$ are viscosity-independent geometric constants. Equations~\ref{Omega_omega} and \ref{v_omega} predict that
\begin{equation}
\Omega  =  R_1 v, \;\;\mbox{with}\; R_1 = \beta/\gamma, \label{purcell} 
\end{equation} 
but underdetermine $(v,\Omega,\omega)$. `Closure' requires experimental input, in the form of the relationship between the torque developed by the motor, $N$, and its angular speed, $\omega_m$, where
\begin{equation}
\omega_m = \Omega + \omega = \left(1+\beta^{-1}\right)\Omega. \label{motor}
\end{equation} 
Measurements have repeatedly shown \cite{Sowa08} that $N(\omega_m)$ displays two regimes, Fig.~\ref{loadline}, which we model as:
\begin{subequations}
\begin{eqnarray}
\omega \leq \omega_m^c  : \;\; N &=& N_0 \label{torque_a}\\
\omega > \omega_m^c  : \;\; N&=&\alpha(\omega_m^{\rm max}-\omega_m) \label{torque_b},
\end{eqnarray}
\end{subequations}
where $\alpha=|\text{d}N/\text{d}\omega_m|=N_0/(\omega_m^{\rm max}-\omega_m^c)$ is the absolute slope of $N(\omega_m)$ when $\omega_m^c < \omega < \omega_m^{\rm max}$. For our purposes later, it is important to realise that Eq.~\ref{motor} implies an equivalent $N(\Omega)$ relation, with associated $\Omega^c$ and $\Omega^{\rm max}$.

Equations~\ref{Omega_omega}, \ref{v_omega} and \ref{torque_a}-\ref{torque_b} completely specify the problem. We can now predict $\Omega$ and $v = \Omega/R_1$, the observables in this work, as functions of solvent viscosity by noting that the motor torque is balanced by the drag torque on the body, i.e.,
\begin{equation}
N= D_0\Omega = \left(\frac{D_0}{1+\beta^{-1}}\right) \omega_m. \label{load}
\end{equation}
Equation~\ref{load} specifies a `load line' that intersects with the motor characteristic curve, Fig.~\ref{loadline},  to determine the `operating condition'. For a prolate ellipsoidal cell body with semi-major and semi-minor axes $a$ and $b$, $D_0=16\pi\eta a b^2 /3$, so that:
\begin{subequations}
\begin{eqnarray}
\omega < \omega_m^c \!\!\!&: &\!\!\! \Omega = \frac{N_0}{D_0}=\left( \frac{3N_0}{16\pi ab^2} \right) \eta^{-1} \label{Omega_a}\\
\omega > \omega_m^c  \!\!\!&: & \!\!\!\Omega=\frac{\alpha^{*}\Omega^{\rm max}}{\alpha^{*}+D_0}=\frac{3\alpha^{*}\Omega^{\rm max} \eta^{-1}}{16\pi ab^2+3\alpha^{*}\eta^{-1}}, \label{Omega_b}
\end{eqnarray}
\end{subequations}
where $\alpha^{*}=|\text{d}N/\text{d}\Omega|=N_0/(\Omega^{\rm max}-\Omega^c)$ is the absolute slope of the $N(\Omega)$ relation (cf. Fig.~\ref{loadline}) in the variable-torque regime.

Recall that BT equated $\Omega \propto \eta^{-1}$ scaling with Newtonian behavior \cite{Berg79}. The above results show that this is true in the constant-torque regime ($\omega < \omega_m^c$) of the motor. Our experiments demonstrate that this is {\it not} the only relevant regime. 

\section{Experimental groundwork: characterising polymers}

SD used `PVP K-90, molecular weight 360,000' \cite{Schneider74}, which, according to current standards \cite{BASF}, has a {\it number averaged} molecular weight of $M_n = 360$~kD, and a weight-average molecular weight of $M_w \approx 10^6$~kD. We show in the online SI that SD's polymer probably has somewhat lower $M_w$ than current PVP~360kD. We used four PVPs (Sigma Aldrich) with stated average molecular weights of $M\sim10$~kD (no K-number given), 40~kD (K-30), 160~kD (K-60) and 360~kD (K-90). Measured low-shear viscosities, which obeyed a molecular weight scaling consistent with good solvent conditions, yielded (see online SI for details) the overlap concentrations \cite{Colby}, $c^* = 0.55 \pm 0.01, 1.4 \pm 0.02, 3.8 \pm 0.1$ and $9.5\pm 0.5$ wt.\% (in order of decreasing $M$), fig.~S2 and Table S1. Static light scattering in water gave $M_w \approx 840$~kD for our PVP360, well within the expected range  \cite{BASF}, and $r_g = 56$~nm, Table S2. We also used Ficoll with $M \sim$~70k and 400k from Sigma Aldrich (Fi70k, Fi400k). 

\section{Results}

We measured the motility of {\it E. coli} in polymer solutions using two new high-throughput methods (see Materials \& Methods and online SI). Differential dynamic microscopy (DDM), which involves correlating Fourier-transformed images in time, delivers, {\it inter alia}, the mean swimming speed $\bar{v}$~\cite{Wilson11,Martinez12}. In dark-field flicker microscopy (DFM),  we average the power spectrum of the flickering dark-field image of individual swimmers to obtain the mean body angular speed, $\bar{\Omega}$. 

Cells suspended in a phosphate motility buffer were mixed with polymer solution in buffer to reach final desired concentrations, and loaded into sealed capillaries for DDM and DFM. The concentrations of cells were low enough to avoid any cell-cell interaction, including polymer-induced `depletion' aggregation \cite{Schwarz12} -- the absence of the latter being confirmed by microscopy. Separate experiments confirmed that oxygen depletion is negligible over the duration of the measurements. 

\begin{figure}[h]
	\vspace*{-.5cm}
	\begin{center}
	\includegraphics[width=3.25in]{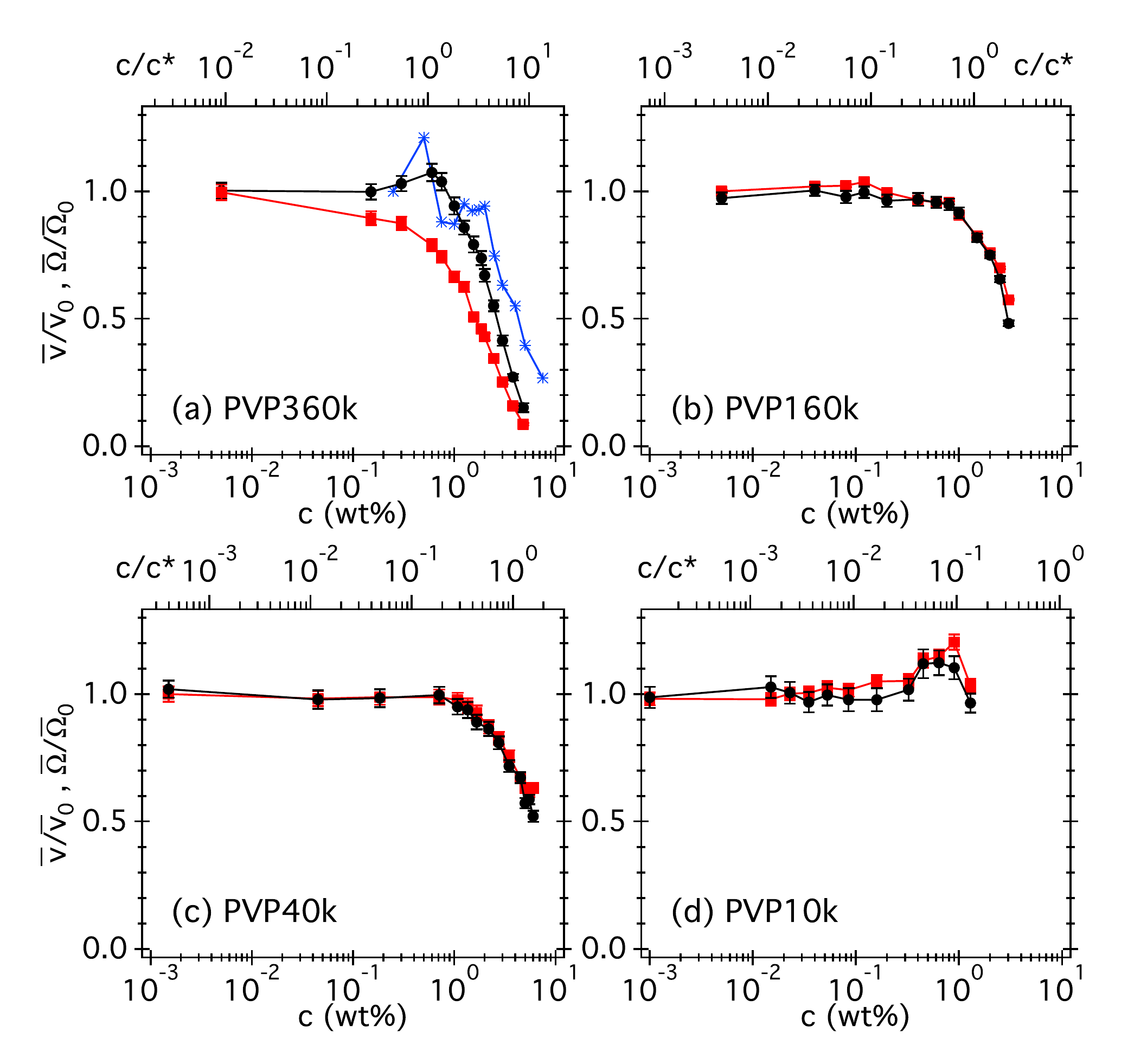}
	\caption{Normalised swimming speed $\bar{v}/\bar{v}_0$ (black circles) and body angular speed $\bar\Omega/\bar\Omega_0$ (red squares) {\it vs.} dialysed PVP concentration (in weight percent) at four molecular weights, with $\bar{v}_0 \approx 15 \mu\text{m/s}$ and $\bar\Omega_0\approx 20\text{Hz}$. The blue stars in (a) are the swimming speeds from SD \cite{Schneider74} normalised to the values at their lowest polymer concentration.}
	\label{fig:normalised_cp}
	\end{center}
\end{figure}

\subsection{Native polymer}
The measured $\bar{v}(c)$ curves for all four PVP (Fig.~S3) and Ficoll (Fig.~S4) solutions are all non-monotonic. The peak we see in PVP~360kD (Fig.~S3) is somewhat reminiscent of SD's observation \cite{Schneider74} for {\it E.~coli} (see also Fig.~2a). Interestingly, all $\bar{\Omega}(c)$ are also non-monotonic except for PVP~360kD (Fig.~S3).

\subsection{Dialysed polymers}
The initial rise in $\bar{v}$ and $\bar{\Omega}$ upon addition of native polymers (Figs.~S3, S4) are somewhat reminiscent of the way swimming speed of {\it E. coli} rises upon adding small-molecule carbon sources (see the example of glycerol in Fig.~S5), which cells take up and metabolise to increase the proton motive force. PVP is highly efficient in complexing with various small molecules \cite{BASF}. We therefore cleaned the as-bought, native polymers by repeated dialysis using membranes that should remove low-molecular-weight impurities (see Materials \& Methods), and then repeated the $\bar v(c)$ and $\bar \Omega (c)$ measurements, Fig.~\ref{fig:normalised_cp}, now reported in normalised form, $\bar{v}/\bar{v}_0$ and $\bar{\Omega}/\bar{\Omega}_0$, where $v_0$ and $\Omega_0$ values at $c = 0$ (buffer).  

The prominent broad peaks or plateaux seen in the data for native PVP40k and PVP160k have disappeared. (The same is true for Fi70k and Fi400k, Fig.~S6.)  A small `bump' (barely one error bar high) in the data for PVP10k remains. Given the flatness of the data in PVP40k and PVP160k, we believe that the residual peak in PVP10k, whose coils have higher surface to volume ratio, is due to insufficient cleaning. On the other hand, a small peak ($\lesssim 10\%$~increase) in $\bar v(c)/\bar{v}_0$ remains for PVP~360kD. For now, what most obviously distinguishes the PVP~360kD from the other three polymers is that the normalised $\bar{v}(c)$ and $\bar{\Omega}(c)$ for the latter coincide over the whole $c$ range, while for PVP~360kD they diverge from each other at all but the lowest $c$. 

\begin{figure}[h]
	\vspace*{-.5cm}
	\begin{center}
	\includegraphics[width=3in]{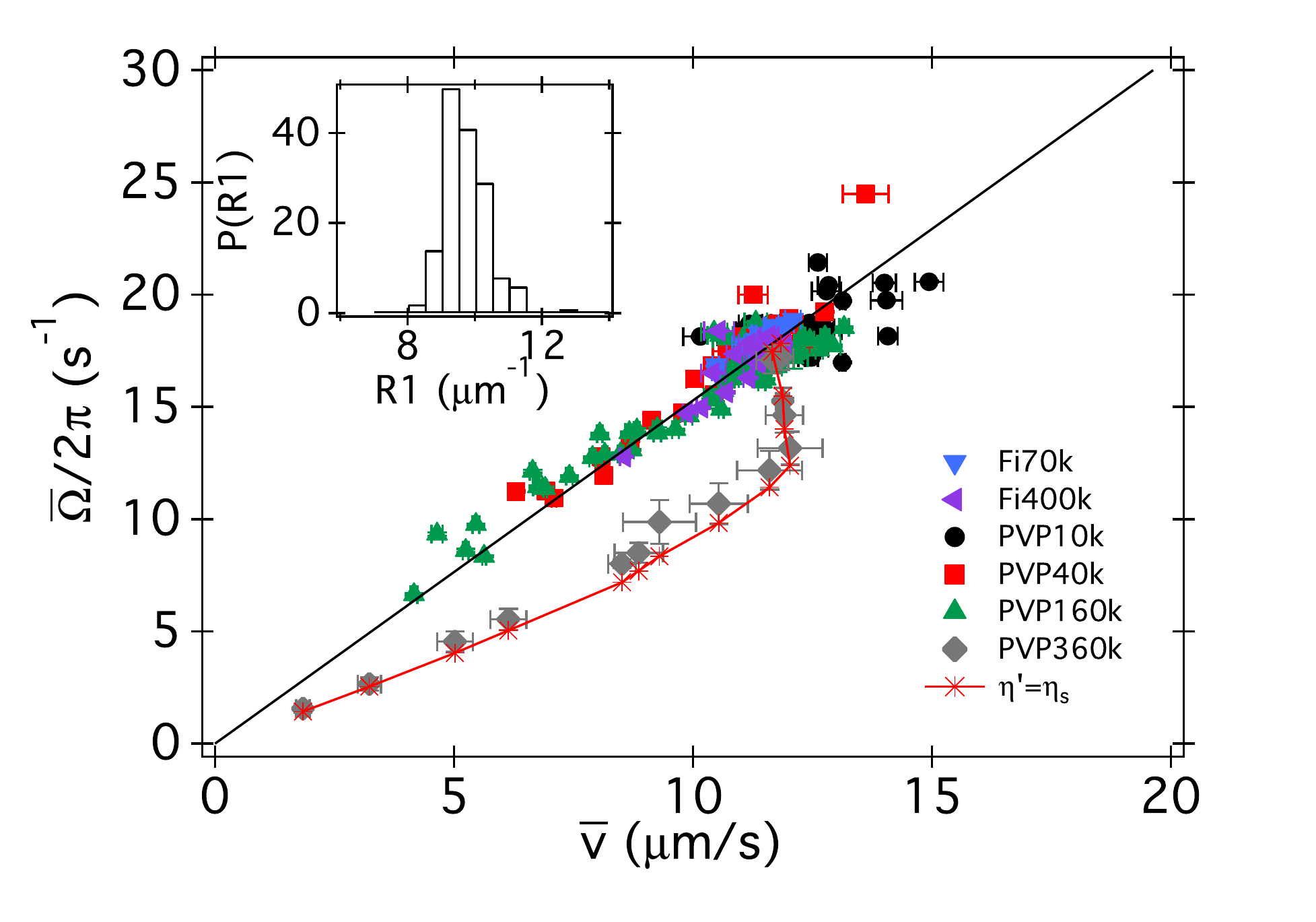}
	\caption{Mean rotational frequency $\bar{\Omega}/2\pi$ versus swimming speed $\bar{v}$ for dialysed PVP and Ficoll solutions at molecular weights as indicated (there are $\geq 2$ datasets per each PVP and one dataset for each Ficoll). The line is a linear fit to all data (except PVP~360kD), giving $R_1= 9.6 \pm 0.1~\mathrm{\mu m}^{-1}$ in Eq.~\ref{purcell}. The inset shows the probability distribution of $R_1$ for all data sets except PVP~360kD. The diamonds are for PVP~360kD averaged over two datasets with the errors bars being standard deviations. The stars linked by the full curve is the predicted $\Omega(v)$ for PVP~360kD, according to a model in which the body experiences the full low-shear viscosity of the polymer solution, and the flagella experiences the viscosity of pure buffer, $\eta_s$.}
	\label{fig:v_body}
	\end{center}
\end{figure}

\subsection{Newtonian propulsion} To observe $\bar v(c)/\bar v_0 = \bar \Omega(c)/\bar\Omega_0$, Fig.~\ref{fig:normalised_cp}(b)-(d), we require $\bar{v} \propto \bar\Omega$, i.e. that Eq.~\ref{purcell} should be valid. This is directly confirmed by Fig.~\ref{fig:v_body}: data for PVP10k, 40k and 160k collapse onto a single master proportionality at all concentrations. Data for two dialysed Ficolls also fall on the same master line. The good data collapse shows that there is only very limited sample to sample variation in the average body and flagellar geometry, which are the sole determinants of $R_1$ in Eq.~\ref{purcell}. The slope of the  line fitted to all the data gives $R_1\approx 9.6 \,\mu \text{m}^{-1}$ (cf. $\approx 7 \,\mu$m$^{-1}$ in \cite{Wu}). The constancy of the ratio $R_1 = \bar{\Omega}/\bar{v}$ is also be seen from the strongly-peaked distribution of this quantity calculated from all individual pairs of $\bar{v}(c)$ and $\bar\Omega(c)$ values except those for PVP~360kD (inset, Fig.~\ref{fig:v_body}).  Physically, $R_1$ is an `inverse cell body processivity', i.e. on average a bacterium swims forward a distance $R_1^{-1} \approx 0.1\, \mu$m per body revolution. 

\begin{figure}[h]
	\begin{center}
	\includegraphics[width=3in]{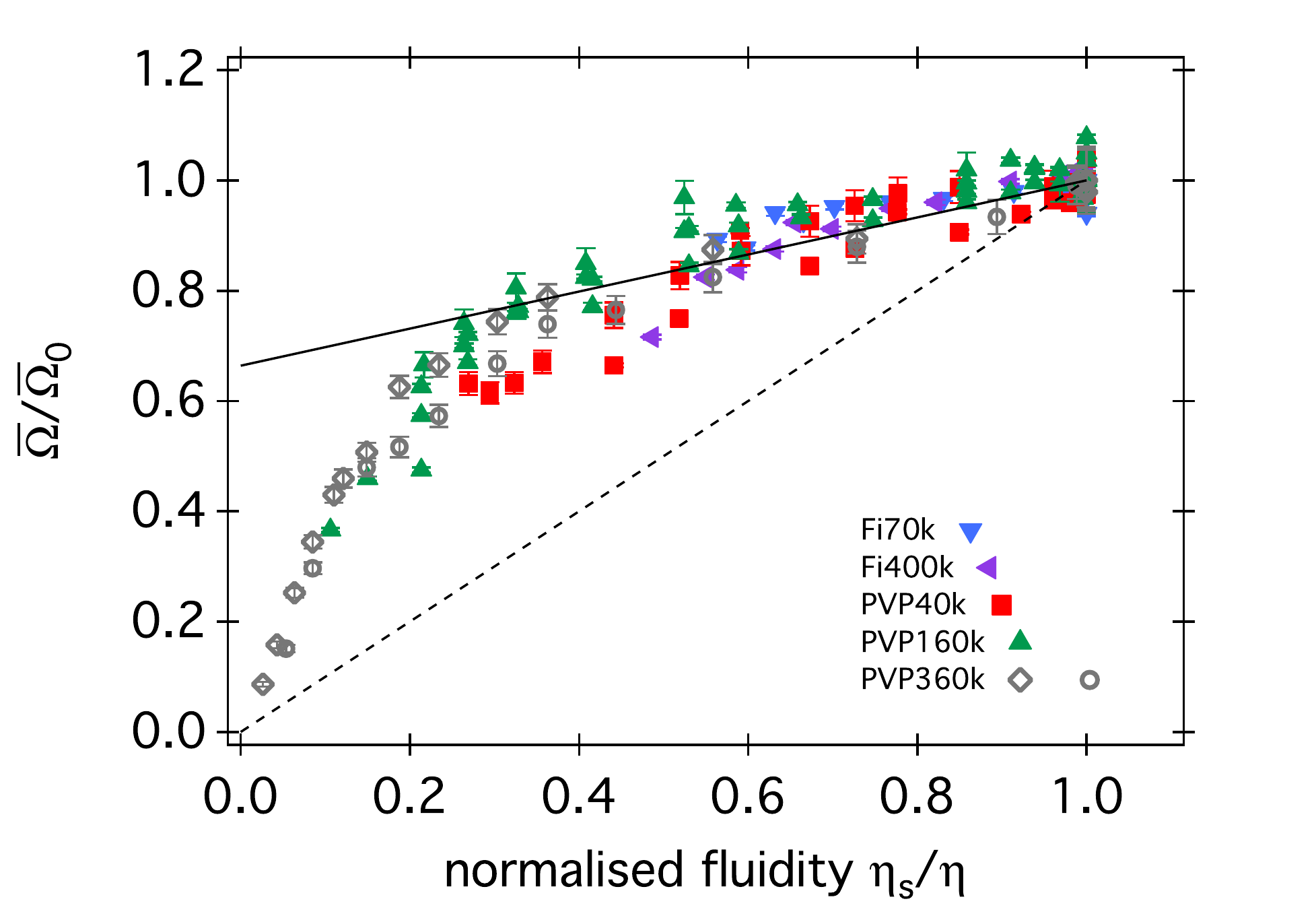}
	\caption{Relative rotational body speed $\bar{\Omega}/\bar{\Omega_0}$ versus fluidity ($1/\eta$), normalised to the fluidity of the motility buffer (c=0), for all polymer solutions we studied. Full and dashed lines are those used by BT to summarise their MC and Ficoll data respectively. Note that BT's MC data spanned a smaller interval ($0.2 \lesssim \eta_s/\eta< 1$) than ours.}
	\label{fig:body_fluidity}
	\end{center}
\end{figure}

The implication of Fig.~\ref{fig:v_body} is that swimming {\it E. coli} sees all our polymer solutions except PVP~360kD as Newtonian fluids.  Interestingly, BT cited the proportionality between $\Omega$ and $\eta^{-1}$ (rather than $\Omega$ and $v$) as evidence of Newtonian behavior in Ficoll. We show the dependence of body rotation speed normalised by its value at no added polymer, $\bar \Omega / \bar \Omega_0$, on the normalised fluidity $\eta_s/\eta$ (where $\eta_s$ is the viscosity of the solvent, i.e. buffer) for our four PVPs and two Ficolls in Fig.~\ref{fig:body_fluidity}, together with the lines used by BT to summarise their MC and Ficoll data. Our data and BT's MC results (which span $0.2 \lesssim \eta_s/\eta< 1$) cluster around a single master curve, which, however, is not a simple proportionality. Equations~\ref{Omega_a} and \ref{Omega_b} together predict such non-linear data collapse, provided that the cell body geometry, $(a,b)$, and the motor characteristics, $(N_0, \Omega^c, \Omega^{\rm max})$, remain constant between data sets. The larger data scatter in Fig.~\ref{fig:body_fluidity} compared to Fig.~\ref{fig:v_body} suggests somewhat larger variations in motor characteristics than in geometry between samples.\footnote{Note, however, that this refers to the fictitious `effective motor' powering the single `effective flagellum' in Purcell's {\it E. coli} model, so that in reality, the variability may reflect differing number and spatial distribution of flagella as much as individual motor characteristics.}

\begin{figure}[h]
	\vspace*{-.5cm}
	\begin{center}
	\includegraphics[width=3in]{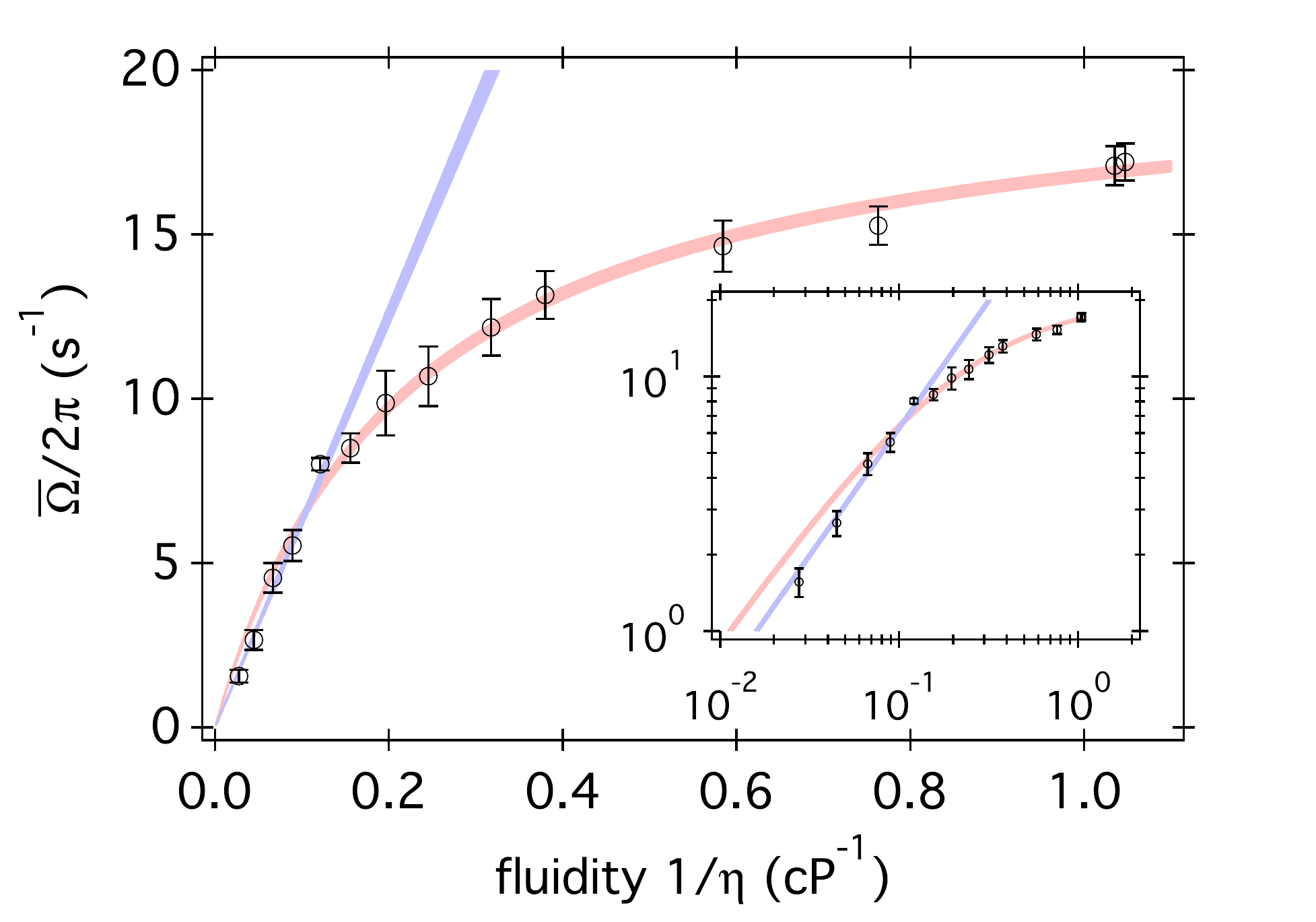}
	\caption{Body rotation frequency versus fluidity averaged over all PVP~360kD data sets. Blue line: fitting the constant-torque result, Eq.~\ref{Omega_a}, in the range $0\lesssim \eta \lesssim 0.15~\text{cP}^{-1}$. Pink curve: fitting the linear-torque result, Eq.~\ref{Omega_b}, in the range $\eta \gtrsim 0.2~\text{cP}^{-1}$. The thickness of the line/curve indicates uncertainties associated with choosing the boundary between the two kinds of behavior. Inset: a log-log plot to show that Eq.~\ref{Omega_b} alone does not fit the data.}
	\label{fig:W_vs_fluidity}
	\end{center}
\end{figure}

Of all the polymers contributing to Fig.~\ref{fig:body_fluidity}, PVP~360kD gave the most extensive coverage over the whole range of fluidity\footnote{To reach lower fluidity, or higher viscosity, required progressively more polymer (by mass). To recover enough polymer after dialysis becomes more challenging as the molecular weight decreases.}, Fig.~\ref{fig:W_vs_fluidity}. Equations~\ref{Omega_a} and \ref{Omega_b} apply to the low and high fluidity regimes of these data respectively. Equation~\ref{Omega_a} depends on a single motor parameter, $N_0$, and predicts a strict proportionality. Our lowest fluidity data points suggest that at the highest polymer concentrations reached, we are indeed operating in this regime. Using $a = 1.2\,\mathrm{\mu m}$ and $b = 0.43\,\mathrm{\mu m}$ (average values from microscopy) to fit Eq.~\ref{Omega_a} to the lowest fluidity data gives $N_0 = 1450 \pm 50$~pN.nm, Fig.~\ref{fig:W_vs_fluidity} (blue), which agrees well with previously measured stall torque \cite{Sowa08}.

The majority of the data away from the lowest fluidities are clearly non-linear, and need to be fitted with Eq.~\ref{Omega_b}. Doing so with the above value of $N_0$ gives $\bar{\Omega}^c/2\pi = 6.6 \pm 0.5$~s$^{-1}$ and $\bar{\Omega}^{\rm max} /2\pi = 20.5 \pm 0.5$~s$^{-1}$, Fig.~\ref{fig:W_vs_fluidity} (pink). Given that $\Omega \propto \omega_m$, Eq.~\ref{motor}, we expect $\Omega^{\rm max}/\Omega^c = \omega_m^{\rm max}/\omega_m^c$. Our ratio of $\bar\Omega^{\rm max}/\bar\Omega^c  \approx 3.1$ compares reasonably with $\omega_m^{\rm max}/\omega_m^c \approx 2.3$ for a different strain of {\it E. coli} at same temperature ($22^\circ$C) \cite{Sowa08}.

Thus, Eqs.~\ref{Omega_a} and \ref{Omega_b} give a reasonable account of the data in Fig.~\ref{fig:W_vs_fluidity}. We conclude that PVP~360kD solution is Newtonian as far as body rotation is concerned.

\subsection{Non-Newtonian effects and flagella nano-rheology} 
Given the above conclusion, the non-linear $\bar\Omega(\bar v)$ for PVP~360kD,  Fig.~\ref{fig:v_body}, suggests a non-Newtonian response at the flagellum. In a minimal model, the flagellum `sees' a different viscosity, $\eta^\prime(c)$, than the cell body, which simply experiences the low-shear viscosity of the polymer solution, $\eta(c)$. Making explicit the viscosity dependence of the resistive coefficients in Eqs.~\ref{bodyFN} and \ref{helixFN} by writing $A = \hat a\eta$, etc., force and torque balance now read:
\begin{eqnarray}\label{theanswer}
\eta\,\hat{a}_0\,v = \eta'\left(-\hat{a}\,v + \hat{b}\,\omega \right), \\
\eta\,\hat{d}_0\,\Omega =\eta'\left(-\hat{b}\,v + \hat{d}\,\omega\right).
\end{eqnarray}
Solving these gives
\begin{equation}
\frac{\Omega}{v} = \frac{\hat{d}\left[\left(\frac{\eta}{\eta'}\right)\hat{a}_0+\hat{a}\right]-\hat{b}^2}{\left(\frac{\eta}{\eta'}\right)\hat{d}_0 \hat{b}}.
\end{equation}

\begin{figure}[h]
	\vspace*{-.5cm}
	\begin{center}
	\includegraphics[width=3in]{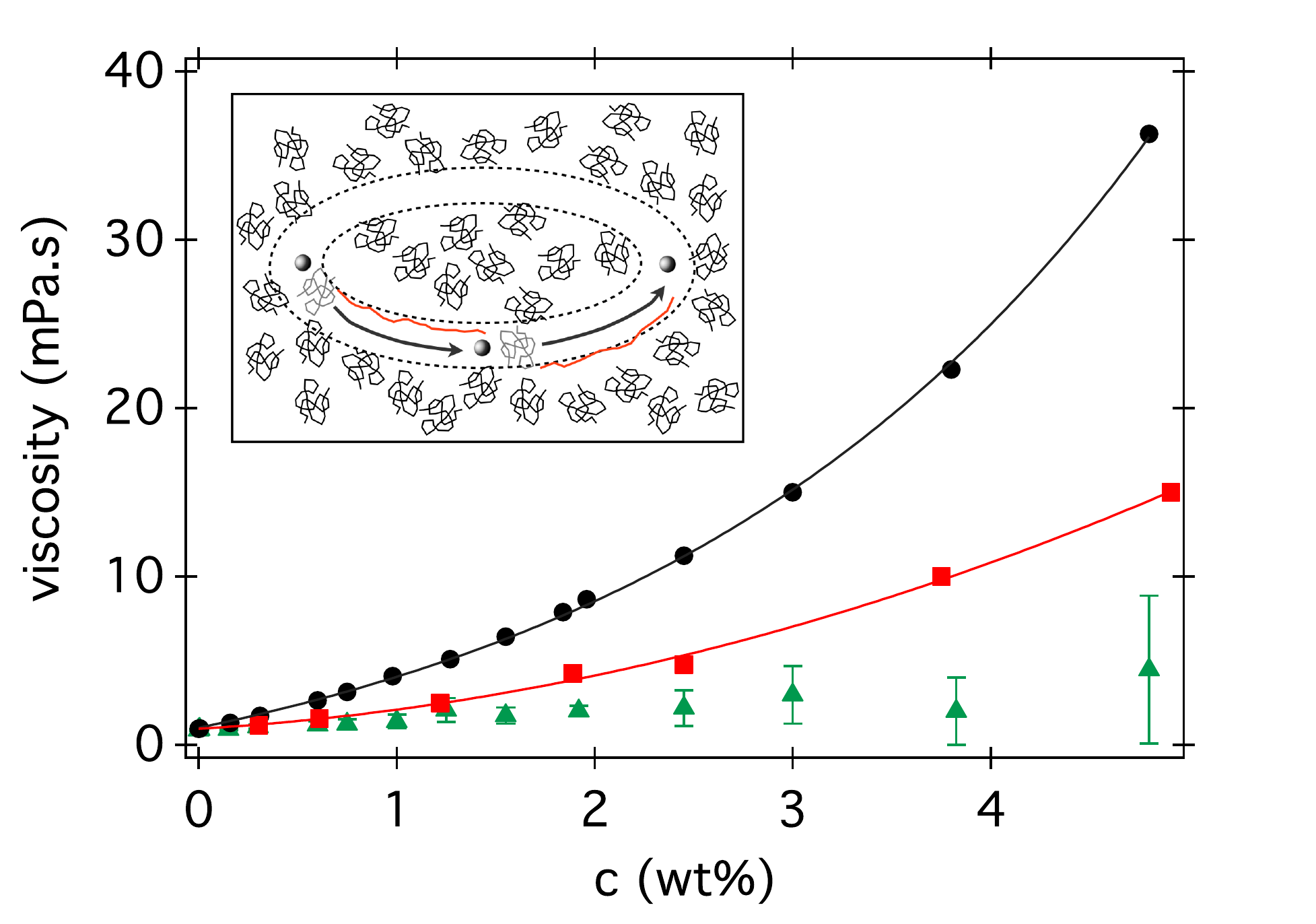}
	\caption{Viscosities of PVP~360kD solutions: low-shear values from rheometry ($\bullet$); micro-rheology data obtained using 980~nm beads at $10^4\,\mathrm{Hz}$ ({\color{red} {\tiny$\blacksquare$}});  $\eta^\prime$ deduced from swimming data  ({\color{forestgreen(web)} {\footnotesize$\blacktriangle$}}). Lines are best fits (see online SI). Inset: schematic showing three snapshots of a section of a flagellum (sphere, $\approx 40$~nm) cutting through a solution polymer coils ($\approx 120$~nm) (with a circular path). Coils, which are initially in the path of the flagellum section (grey), become stretched out (red), leaving a coil-sized channel of solvent}.
	\label{nanorheology}
	\end{center}
\end{figure}

Equation~\ref{theanswer} makes an interesting prediction. If we take $\eta^\prime(c) = \eta_s$ and use previously-quoted flagellum dimensions for {\it E. coli} \cite{Wu} to calculate $(\hat a_0, \hat d_0; \hat a, \hat b, \hat d)$,  it predicts nearly perfectly the observed non-linear $\bar\Omega(\bar v)$ relationship for PVP~360kD, Fig.~\ref{fig:v_body}. Details are given in the online SI, where we also predict the observed peak in $\bar{v}(c)$, Fig.~\ref{fig:normalised_cp} (Fig.~S7). To check consistency, we proceed in reverse and  treat the flagellum as a nano-rheometer. Given the measured $\bar{v}(c)$ in PVP~360kD, we deduce the viscosity seen by the flagellum, $\eta^\prime(c)$, at shear rate $\dot\gamma \approx 10^4\,\mathrm{s^{-1}}$ (details in online SI), Fig.~\ref{nanorheology}, where we also show the low-shear viscosity of PVP~360kD solutions measured using conventional rheometry. Indeed, over most of the concentration range, we find $\eta^\prime \approx \eta_s$. (Note that the highest $c$ data points are subject to large uncertainties associated with measuring very low swimming speeds.) Thus, our data are consistent with the flagellum `seeing' essentially just the viscosity of the pure solvent (buffer). Macroscopically, this corresponds to extreme shear thinning. Is this a reasonable interpretation?

For a helical flagellum of thickness $d$ and diameter $D$ rotating at angular frequency $\omega$, the local shear rate is $\dot\gamma_f\sim\omega D/d$ (we neglect translation  because $v \ll \omega D$). For an \emph{E. coli} flagellar bundle, $d\approx 40$ nm, $D\approx 550$ nm and $\omega \approx 2\pi\times 115$ rad/s \cite{Wu}, giving $\dot\gamma_f\lesssim10^4$ s$^{-1}$ in the vicinity of the flagellum. The Zimm relaxation time of a polymer coil is $\tau_Z\sim~4\pi\eta_s~r_g^3 /k_B T$, where $k_B T$ is the thermal energy. Using $\eta_s=10^{-3}$ Pa$\cdot$s, $r_g\sim 60$ nm, we find $\tau_Z\sim1$~ms for our PVP~360kD at room temperature. Since $\Omega^{-1} \gg \tau_Z$, the cell body does not perturb significantly the polymer conformation. However, $\dot\gamma_f^{-1} \sim 0.1\tau_Z$, so that the polymer may be expected to shear thin in both dilute ($c < c^*$) \cite{Larson} and semi-dilute ($c \gtrsim c^*$) \cite{Gompper} solutions. Low-shear rate data collected using rheometry and high-frequency microrheological data collected using $1 \mu$m beads and interpreted using the Cox-Merz rule \cite{CoxMerz58} (see online SI for details) show that there is indeed significant shear thinning of our PVP~360kD polymer, Figs.~\ref{nanorheology} and S8, though not as extreme as thinning down to $\eta_s$. 

Anisotropic elastic stresses \cite{LaugaSheet,ShelleySheet,FuFilament,BreuerHelix} and shear thinning \cite{Smith2013,Lauga13} have been proposed before as a possible cause of non-Newtonian effects in biological swimmers. However, in the usual sense, these are continuum concepts arising from experiments on the mm (rheometry) or $\mu$m (microrheology) scale. Neither is obviously  applicable to a $\sim 40$~nm segment of flagellum moving through somewhat larger polymer coils ($r_g \sim 60$~nm). One of the very few explorations of the `probe $\approx$ polymer size' regime to date found a highly-non-linear time dependent response with a shear-thinning steady state that matched bulk rheometry data, albeit with quite stiff polymers ($\lambda$-phage DNA) \cite{Cribb}. The relevant physics may be similar to, but perhaps more complex than, the active microrheology of colloidal suspensions using `probes' that are approximately the same size as the colloids \cite{SquiresBrady}. A qualitative picture Fig.~6 (inset) may be as follows. A section of the flagellum travelling at $\omega D/2 \sim 200 \mu$m/s takes $\sim 0.3\tau_Z$ to traverse $\sim r_g$. Thus, polymer coils in its vicinity are strongly stretched as quasi-stationary objects and the flagellum effectively carves out' a $\sim 2 r_g$-wide channel practically void of polymers. Each flagellum section revisits approximately the same spatial location with a period $2\pi/\omega \sim 10$~ms (because the translation per turn is low). Although this time is larger than the single chain relaxation time $\tau_Z$, the time required for collective relaxation and diffusion of a large number of strongly stretched polymer chains is significantly larger than that. Effectively, then, the flagellum moves inside a channel with viscosity $\rightarrow \eta_s$. Moreover, under strong local elongation of the kind we have suggested, it is also possible that polymers may break \cite{Caruso}. This is a separate, but related, mechanism for change in mechanical properties of the solution around the flagellum. 

We note that previous experiments of {\it E. coli} swimming in MC \cite{Berg79,Darnton} used polymers and worked in concentration regimes where shear thinning effects are insignificant. There is already indirect evidence of this in Fig.~\ref{fig:body_fluidity}, where data from BT \cite{Berg79} collapse onto a Newtonian master curve for $\bar\Omega(\eta^{-1})$. More directly, these previous studies used methyl celluloses with `viscosity grade' around 4000 cP in the range 0-0.3 wt.\% \cite{Berg79} and at 0.18 wt.\% \cite{Darnton}. The shear thinning of such polymers have been measured (polymer AM4 in \cite{Edelby})  and fitted to a power law: $\eta \sim \dot\gamma^{n-1}$; at $c = 0.25$~wt.\% and 0.5~wt.\%, $n = 1.00$ and 0.961 respectively. Thus, at the concentrations used before~\cite{Berg79,Darnton}, shear thinning is very weak or absent, and the solutions behave as Newtonian.

\section{Summary and Conclusions}

We have measured the average swimming speed and cell body rotation rate in populations of {\it E. coli} bacteria swimming in different concentrations of solutions of the linear polymer PVP (nominal molecular weights 10kD, 40kD, 160kD and 360kD, these probably being number-averaged values) and the branched polymer Ficoll (70kD and 400kD). We dialysed each polymer to remove small-molecular impurities that can be metabolised by the cells to increase their swimming speed. The collapse of data for all polymers except PVP~360kD onto a single proportionality relationship between swimming speed and body rotation rate, $\bar\Omega(\bar v)$, Fig.~\ref{fig:v_body}, demonstrates that these solutions behave as Newtonian fluids as far as {\it E. coli} propulsion is concerned. 

Significant non-linearities in $\bar\Omega(\bar v)$ were found for {\it E. coli} swimming in PVP~360kD solutions. Further analysis showed that the motion of the cell body remained Newtonian: the measured $\bar\Omega(\eta^{-1})$ can be fitted to results derived from Newtonian hydrodynamics (Eqs.~\ref{Omega_a} and \ref{Omega_b}), Fig.~\ref{fig:W_vs_fluidity}. Thus there must be non-Newtonian effects at the flagellum. The observed deviations from Newtonian behavior can be quantitatively accounted for by a simple model in which the flagellum `sees' the viscosity of pure buffer. This is consistent with significant shear thinning observed at the micron level in PVP~360kD solutions using microrheology, although we suggest that molecular effects must be taken into account because the polymer and flagellum filament have similar, nanometric, dimensions. Note that the effects we are considering, which arise from high shear rates, are absent from experiments using macroscopic helices as models for viscoelastic flagella propulsion \cite{BreuerHelix}.

Shear thinning is not the only possible effect in the vicinity of a flagellum creating local deformation rates of $\sim 10^4 \mathrm{s^{-1}}$. Higher molecular weight polymers that are more viscoelastic than PVP~360kD will show significant elastic effects.  Interestingly, it is known that double-stranded DNA could be cut at a significant rate at $\dot\gamma \sim 10^4$s$^{-1}$ \cite{Zimm77}. An {\it E. coli} swimming through high molecular weight DNA solution should therefore leave behind a trail of smaller DNA and therefore of lower-viscosity solution, making it easier for another bacterium to swim in the wake. This may have important biomedical implications: the mucosal lining of normal mammalian gastrointestinal tracks and of diseased lungs can contain significant amounts of extracellular DNA.
Exploration of these issues will be the next step in seeking a complete understanding of flagellated bacterial motility in polymeric solution. 

\section{Materials and Methods}

\subsection{Cells}
We cultured K12-derived wild-type {\em E. coli} strain AB1157 as detailed before \cite{Wilson11,Martinez12}. Briefly, overnight cultures were grown in Luria-Bertani Broth (LB) using an orbital shaker at 30$^\circ$C and 200~rpm. A fresh culture was inoculated as 1:100 dilution of overnight grown cells in 35ml tryptone broth (TB) and grown for 4~h (to late exponential phase). Cells were washed three times with motility buffer (MB, pH = 7.0, 6.2 mM K$_2$HPO$_4$, 3.8 mM KH$_2$PO$_4$, 67 mM NaCl and 0.1 mM EDTA) by careful filtration (0.45~$\mu$m HATF filter, Millipore) to minimize flagellar damage and resuspended in MB to variable cell concentrations.

\subsection{Polymers} 
\emph{Native}: PVP and Ficoll from Sigma-Aldrich were used at four (10k, 40k, 160k, 360k) and two (70k and 400k) nominal molecular weights respectively. Polymer stock solutions were prepared and diluted with MB. \emph{Dyalisis}: the polymer stock solutions were dialyzed in tubes with 14 mm diameter and 12 kDa cut-off (Medicell International Ltd) against double-distilled water. The dialysis was performed over 10 days with daily exchange of the water. The final polymer concentration was determined by measuring the weight loss of a sample during drying in an oven at $55^\circ$C and subsequent vacuum treatment for 6h. Polymer solutions at several concentration were prepared by dilution using MB.

\subsection{Motility measurement} $~$ Bacterial suspensions were gently mixed with the polymer solutions to a final cell density of $\approx 5\times10^8$~cells/ml. A $\approx~400~\mu\text{m}$ deep flat glass sample cell was filled with $\approx~150~\mu\text{l}$ of suspension and sealed with Vaseline to prevent drift. Immediately after, two movies, one in phase-contrast illumination ($\approx$~40s-long, Nikon Plan Fluor 10$\times$Ph1 objective, NA~=~0.3, Ph1 phase-contrast illumination plate at 100 frame per second and $500^2$ pixels) and one in dark-field illumination ($\sim$10s-long, Nikon Plan Fluor 10$\times$Ph1 objective, NA~=~0.3, Ph3 phase-contrast illumination plate, either 500 or 1000 frame per second, $500^2$ pixels) were consecutively recorded on an inverted microscope (Nikon TE300 Eclipse) with a Mikrotron high-speed camera (MC 1362) and frame grabber (Inspecta 5, 1~Gb memory) at room temperature ($22\pm1^{\circ}$C). We image at $100~\mu\text{m}$ away from the bottom of the capillary to avoid any interaction with the glass wall.

We measured the swimming speed from the phase contrast movies using the method of different dynamic microscopy (DDM) as detailed before \cite{Wilson11,Martinez12}. The dark field movies were analysed to measure the body rotation speed using the method of dark-field flicker microscopy (DFM), in which we Fourier transform the power spectrum of the flickering image of individual cells, and identify the lowest frequency peak in the average power spectrum (Fig.~S9) as the body rotation frequency as in previous work \cite{Wu,Lowe87}; the difference here is that DFM is a high-throughput method (see online SI).

\subsection{Rheology}We measured the low-shear viscosity $\eta$, of polymer solutions using a TA Instruments AR2000 rheometer in  cone-plate geometry (60~cm, 0.5$^{\circ}$). 
Passive micro-rheology was performed using diffusing wave spectroscopy in transmission geometry with 5 mm thick glass cuvettes. The used set-up (LS Instruments, Switzerland) uses an analysis of the measured mean square displacement (MSD) of tracer particles as detailed before \cite{Mason2000}. Tracer particles (980 nm diameter polystyrene) were added to the samples at 1 wt$\%$ concentration. The transport mean free paths $l^*$ of the samples were determined by comparing the static transmission to a reference sample (polystyrene with 980 nm diameter at 1 wt$\%$ in water). The shear rate dependent viscosity $\eta$ was obtained from the frequency dependent storage and loss moduli using the Cox-Merz rule \cite{CoxMerz58}.

\begin{acknowledgments}
The work was funded by the UK EPSRC (D071070/1, EP/I004262/1, EP/J007404/1), EU-FP7-PEOPLE (PIIF-GA-2010-276190), EU-FP7-ESMI (262348), ERC (AdG 340877 PHYSAPS), and SNF (PBFRP2-127867).
\end{acknowledgments}

\end{article}

\begin{center}
{\Large Supporting Information}
\end{center}
\vspace{0.5cm}

\begin{article}
\setcounter{footnote}{0}
\setcounter{equation}{0}
\setcounter{figure}{0}
\renewcommand{\thefigure}{S\arabic{figure}}

\section{Predictions from Magariyama and Kudo \cite{Magariyama02}}
\begin{figure*}[h]
	\begin{center}
	\includegraphics[width=3.25in]{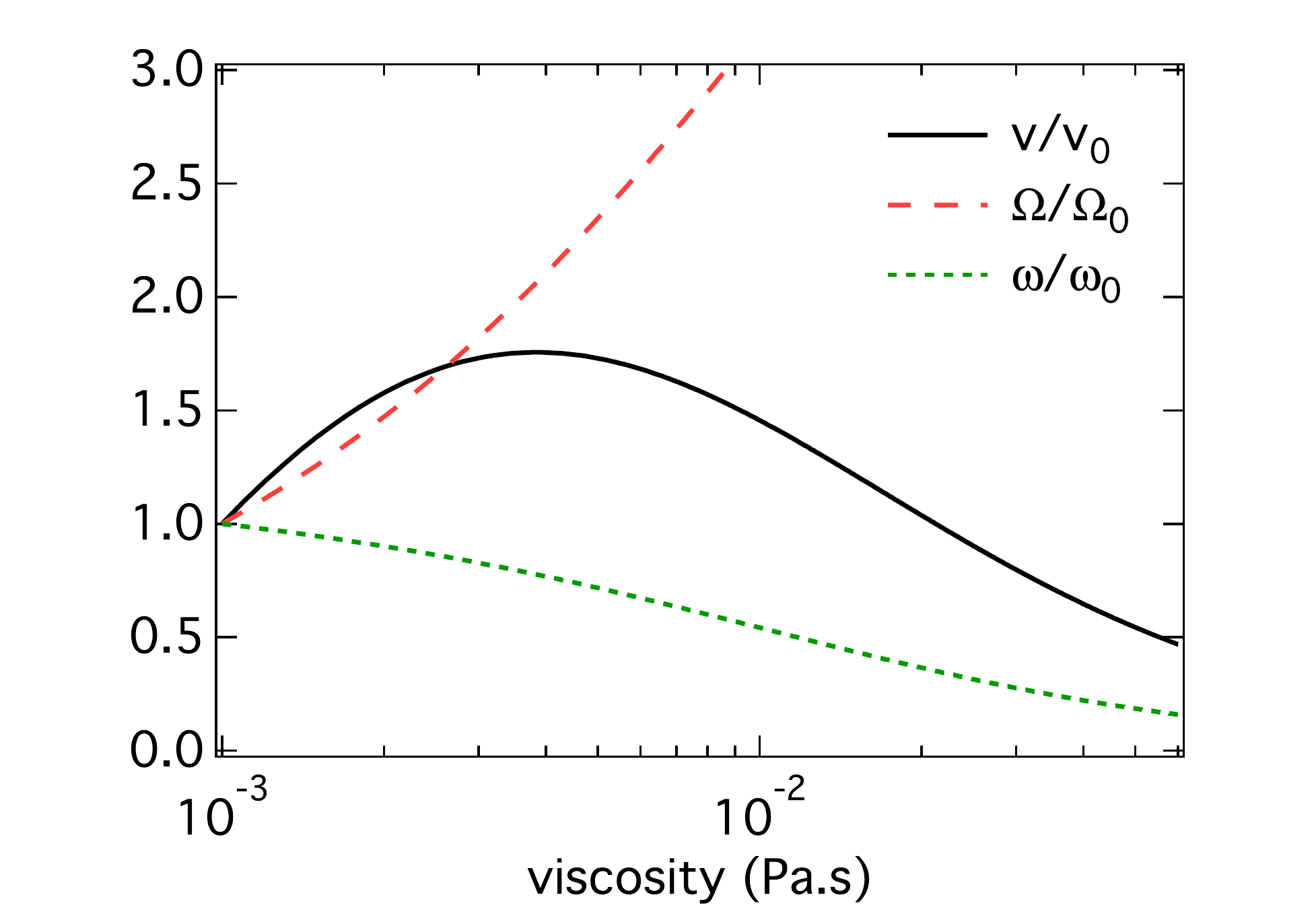}
	\caption{Normalised swimming speed $v/v_0$, body rotational speed $\Omega/\Omega_0$, and flagella rotational speed $\omega/\omega_0$ versus viscosity according to Magariyama $\&$ Kudo \cite{Magariyama02}.}
	\label{theory}
	\end{center}
\end{figure*}

We have calculated the predicted dependence of swimming velocity on polymer concentration using the theory of Magariyama and Kudo \cite{Magariyama02}, Fig.~S\ref{theory}, and show that, as they claimed, there is a peak. However, we also show a calculation that they did not report, namely, the predicted dependence of body rotation frequency as a function of polymer concentration. The latter is a rapidly increasing function, which is clearly unphysical, and contradicts observation by Berg and Turner \cite{Berg79}, as well as data shown here (Fig.~4, main text). 

\section{Characterising the PVP}

\subsection{Intrinsic viscosity and overlap concentration \cite{Colby}}
The viscosity of a polymer solution at low concentrations can be written as a virial expansion of $\eta$ in $c$:
\begin{equation}
\eta=\eta_s \left(1+[\eta]c+k_H[\eta]^2c^2+...\right)
\end{equation}
where $[\eta]$ and $k_H$ are the intrinsic viscosity and the Huggins coefficient respectively, and $\eta_s$ is the viscosity of the solvent, here motility buffer. The linearity at low $c$ can be expressed in two different ways:
\begin{eqnarray}
\frac{\eta-\eta_s}{\eta_s c} & = & [\eta]+k_H [\eta]^2 c \;\;\;\mbox{(Huggins)} \label{eq:huggins}\\
\frac{\text{ln}(\eta/\eta_s)}{c} &= & [\eta]+(k_H-\frac{1}{2})[\eta]^2 c \;\;\;\mbox{(Kraemer)} \label{eq:kraemer}
\end{eqnarray}
These two linear plots should extrapolate to $[\eta]$ at $c = 0$. The intrinsic viscosity measures the volume of a polymer coil normalised by its molecular weight, so that $c^* \approx [\eta]^{-1}$. A modern text names this as the best experimental method for estimating the overlap concentration \cite{Colby}.

\begin{figure*}[h]
	\begin{center}
	\includegraphics[width=6.5in]{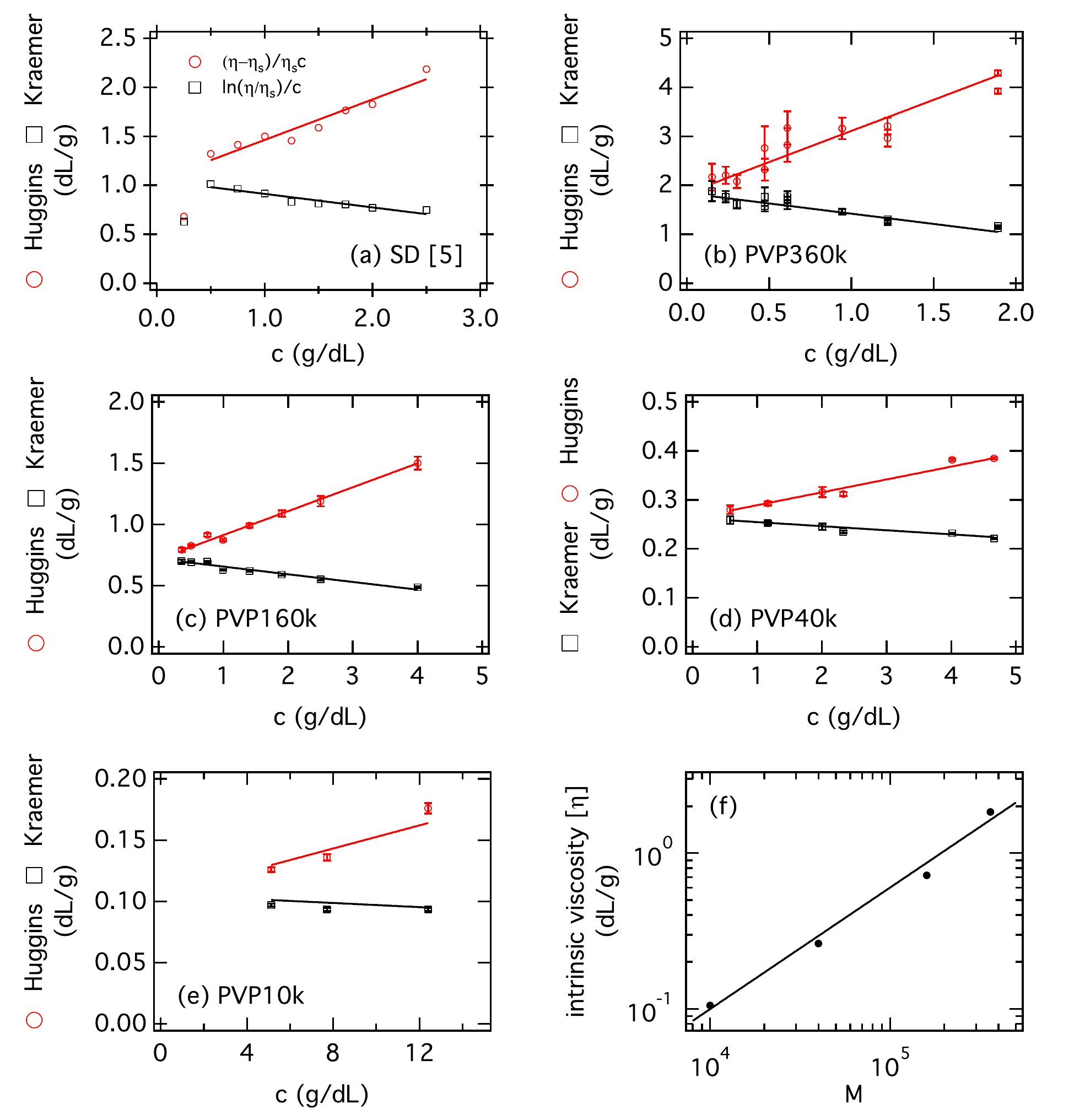}\\
	\caption{(a)-(e) Huggins \& Kraemer representation. (red circles) $(\eta-\eta_s)/\eta_s c$ and (black squares) $\text{ln}(\eta/\eta_s)/c$ versus polymer concentration. Lines are linear fits to the data using Eq.~\ref{eq:huggins} and Eq.~\ref{eq:kraemer} simultaneously. Both quantities should be linear, and extrapolate to a unique intrinsic viscosity $[\eta]$ at $c = 0$. (a) From the PVP viscosity data of Schneider and Doetsch. Discarding the lowest-$c$ point gives $[\eta] = 1.05\pm0.02$. (b)-(e) Our PVP at four different molecular weights. (f) The scaling of intrinsic viscosity, $[\eta]$, with molecular weight, $M$, for our PVPs.}
	\label{scaling}
	\end{center}
\end{figure*}

We first re-graph the $\eta(c)$ data given by Schneider and Doetsch \cite{Schneider74} as Huggins and Kraemer plots, Fig.~S\ref{scaling}(a). (Note for here and below that concentrations in wt\% and g/dL are interchangeable at the sort of concentrations we are considering.) It is clear that their lowest-$c$ data point must be inaccurate. Discarding this point gives the expected linear dependence in both plots, and a uniquely extrapolated value of $[\eta] = 1.055$ at $c = 0$, giving $c^* = 0.95$~g/dL. Reference to our $c^*$ values below suggests that SD's PVP 360~kD has somewhat lower molecular weight than our material with the same label. 

According to current industry standards \cite{BASF}, PVP 360~kD should have viscosities of $\approx 3-5$~mPa.s  and $\approx 300-700$~mPa.s at 1~wt.\% and 10~wt.\% in water respectively.  SD's reported viscosities at 1 and 10~wt.\%, 2.5 and 249~Pa.s respectively, are lower than these values, again consistent with their material having lower molecular weight than our PVP 360~kD.

We have characterised all four PVPs used in this work by measuring their low-shear viscosity in motility buffer as a function of concentration. For K-90 at 1~and 10 wt.\%, we found $\eta \approx 4$ and 370 mPa.s, agreeing well with the published standards \cite{BASF}.  We now graph the measured viscosities of our four PVPs at low concentrations as Huggins and Kraemer plots, Fig.~S\ref{scaling}(b)-(e). In each case, the expected behavior is found; the extrapolated values of $[\eta]$ and the overlap concentrations calculated from these are given in Table~\ref{table:rheodata}. The scaling of $[\eta]$ vs. $M$ is consistent with a power law, Fig.~S\ref{scaling}(f), $[\eta] \sim M^a$ with $a = 0.781$. Since $[\eta] \approx r^3/M$, so $r \sim M^\nu$ with $\nu = (1+a)/3$. We find $\nu = 0.593$, which is consistent with the renormalization group value of $\nu = 0.588$ for a linear polymer in a good solvent. 

\subsection{Coil radii, second virial coefficient and molecular weight \cite{Berne}}
We performed static and dynamic light scattering (SLS $\&$ DLS) experiments to measure the radius of gyration, $R_g$, the molecular weight, $M_w$, the second virial coefficient, $A_2$, and the hydrodynamic radius, $R_h$, of PVP360k in water and motility buffer. $R_h$ was measured by DLS, and $R_g$, $M_w$, and $A_2$ using the Zimm plot of SLS data. Results are summarised in Table \ref{table:zimmdata}. The positive $A_2$ is consistent with our conclusion above that water is a good solvent for PVP. There may be a mild degree of aggregation in motility buffer (larger radii and slightly smaller $A_2$).  

\section{Native Polymer Results}
\begin{figure*}[h]
	\begin{center}
	\includegraphics[width=3.25in]{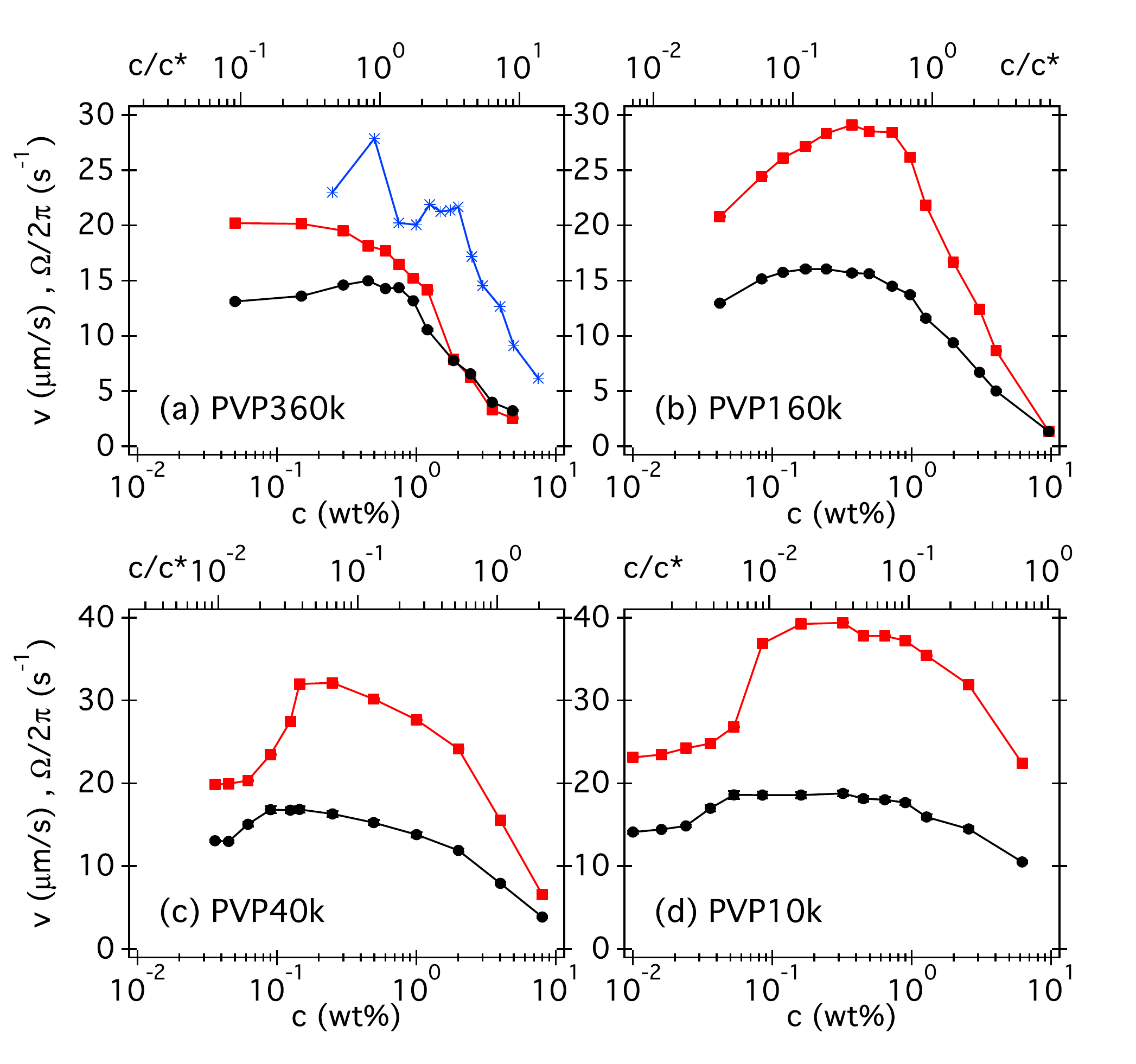}
	\caption{Swimming speed $\bar{v}$ (black circles) and body rotation frequency $\bar{\Omega}/2\pi$ (red squares) of {\it E. coli} {\it vs.}  concentration (in weight percent) of native PVP of four molecular weights. The stars (blue) in (a) are results for swimming speed from SD \protect{\cite{Schneider74}}.}
	\label{fig:U_v_cp}
	\end{center}
\end{figure*}

Figure S\ref{fig:U_v_cp}(a) shows $\bar{v}$ and $\bar{\Omega}$ versus polymer concentration, $c$, for as-bought, or native, PVP360k. While $\bar{\Omega}(c)$ decays monotonically, a peak is observed in $\bar{v}$ at $c \approx0.5\text{wt}\%$, or roughly $c^*$ for this molecular weight. The latter ostensibly reproduces SD's observations \cite{Schneider74} -- their data are also plotted in Fig.~S\ref{fig:U_v_cp}(a). In native PVP160k, Fig.~S\ref{fig:U_v_cp}(b), the peak in $\bar v(c)$ broadens, and now there is a corresponding broad peak in $\bar{\Omega}(c)$ as well. These peaks broaden out into plateaux for native PVP40k and PVP10k, Fig.~S\ref{fig:U_v_cp}(c-d).  We also performed experiment with native Ficoll with manufacturer quoted molecular weights of 70k and 400k, and observed similar non-monotonic, broadly peaked responses in both $\bar v(c)$ and $\bar \Omega (c)$ (Fig.~S4).

\section{The effect of small-molecule energy sources}
Here we show $\bar{v}(c)$ for {\it E. coli} swimming in glycerol solutions of a range of concentrations, Fig.~S\ref{smallmol}. The plot is indeed reminiscent of what is seen for native PVP10k and PVP40k. Indeed, we suggest that the increases at low concentrations in all four polymers have the same origin as the increase observed at low glycerol concentration -- the availability of a small-molecule energy source. The decrease at high glycerol is an osmotic effect (as observed for other small-molecules, e.g. sucrose \cite{Pilizota2013}), while that seen in the 10k, 40k and 160k polymers can be entirely accounted for by low-Re Newtonian hydrodynamics (Polymeric osmotic effects at our concentrations are negligible). 

\section{Dialysed Ficoll results}
Swimming speed and body rotation frequency as a function of concentration is shown for two purified Ficolls in Fig.~S6.

\section{Shear-thinning calculations}
\subsection{Predicting $\Omega(v)$ for flagellum experiencing buffer viscosity}
Here we outline the procedure used to calculate the rotation rate of the cell body for a bacterium swimming in shear-thinning PVP360k solution as a function of the swimming speed. We assume that the flagellum `sees' a viscosity $\eta'$ that is different from the low-shear-rate viscosity $\eta$ experienced by the bacterial body. This is partly motivated by bulk and microrheological measurements, Fig.~6 in the main text, showing that at the shear rates generated by the flagellum, shear thinning can be expected at least down to the $\mu$m scale. Empirically, the low- and high-shear viscosities plotted in Fig.~6 of the main text can be fitted by:
\begin{equation}
\begin{cases}
\eta_{\text{low-shear}} = -5.32 + 6.33\exp{\left(0.39c\right)} & \dot\gamma\rightarrow 0 s^{-1},\\
\eta_{\text{high-shear}} = 0.96 + 0.69c + 0.44c^2 & \dot\gamma=10^4 s^{-1},
\end{cases}
\end{equation}
where $c$ is in wt\% and $\eta$ in mPa$\cdot$s. 

In this two-viscosity model, the force and torque balance equations solve to Eq.~13 in the main text:
\begin{equation}
\frac{\Omega}{v} = \frac{\hat{d}\left[\left(\frac{\eta}{\eta'}\right)\hat{a}_0+\hat{a}\right]-\hat{b}^2}{\left(\frac{\eta}{\eta'}\right)\hat{d}_0 \hat{b}}.
\label{omega_theory}
\end{equation}
The friction coefficients in Eq.~(\ref{omega_theory}) are given by \cite{Wu}
\begin{eqnarray}
&& \hat{a} = k_n L \sin{\psi} \tan{\psi} \left( 1+\gamma \cot^2{\psi} \right), \\
&& \hat{b} = k_n L \frac{\lambda}{2\pi}\sin{\psi} \tan{\psi} \left( 1-\gamma \right), \\ 
&& \hat{d} = k_n L \left(\frac{\lambda}{2\pi}\right)^2 \sin{\psi} \tan{\psi} \left(  1+\gamma \cot^2{\psi} \right), \\
&& \hat{a}_0 = \frac{4\pi b}{\ln{\frac{2b}{a}-\frac{1}{2}}}, \\
&& \hat{d}_0 =  \frac{16\pi}{3}a^2 b,
\end{eqnarray}
where
\begin{eqnarray}
k_n = \frac{8\pi}{2\ln{\frac{c\lambda}{r}}+1}, \\
k_t = \frac{4\pi}{2\ln{\frac{c\lambda}{r}}-1},
\end{eqnarray}
and $\gamma=k_t/k_n$. Here, $L = 7 \mu$m and $\lambda = 2 \mu$m are the total length and pitch of the flagellum, correspondingly, $\psi = 41^{\circ}$ is the angle made by the flagellar filament with the flagellar axis, $r = 20$nm is the estimated radius of a the composite filament in a flagella bundle, and $c=2.4$ is the Lighthill constant. All parameters are taken from a previous experimental paper \cite{Wu}, where this set of parameters were shown to be consistent with the Purcell model. 

Using the measured values of $\bar{\Omega}(c)$ for PVP360k, $\eta'=\eta_s$, $\eta=\eta_{\text{low-shear}}$, and Eq.~(\ref{omega_theory}) is sufficient to calculate the corresponding $\bar{v}(\bar{\Omega})$. Results show good agreement with the measured values (Fig.~3 in the main text), thus predicting a peak in the swimming velocity upon an increase in the viscosity of the polymer solution. For a better illustration, we compare the predicted and measured values of $\bar{v}$ as a function of the viscosity experienced by the body ($\eta_{\text{low-shear}}$) in Fig.~S\ref{v_vs_eta}. Our theory is successful in predicting a peak in the swimming velocity in the right position and of the right shape.

\subsection{Deducing the viscosity the flagellum sees from measurements}
Now we relax our previous assumption that $\eta'$ is equal to the viscosity of the solvent, and use Eq.~(\ref{omega_theory}) to extract the viscosity of the fluid surrounding the flagellar filament. Using the measured values of $\bar{\Omega}(c)$ and $\bar{v}(c)$, Eq.(\ref{omega_theory}) can be solved for $\eta'$. The results are shown in Fig.~6 in the main text. Indeed, for most of the concentration range studied, $\eta^\prime \approx \eta_s$.

\section{Dark-field Flicker Microscopy}
Under dark-field illumination, the image of a swimming bacterium appears to flicker. By calculating the power spectrum of the spatially localised time-dependent intensity fluctuations of low-magnification images of a quantised pixel box (containing $\approx$ 1 cell), and then averaging over all cells in the images, we are able to measure the body rotational frequency $\Omega/2\pi$ averaged over $\sim 10^4$ cells based on a $\lesssim 10$ s movie. This method is similar to what was done by Lowe et al. \cite{Lowe1987}, who measured the power spectrum of single swimming cells. However, here we use low-magnification dark-field imaging which allows high-throughput measurement of $\Omega/2\pi$.

Dark-field movies were recorded (Nikon Plan Fluor 10$\times$Ph1 objective, NA = 0.3, Ph3 phase-contrast illumination plate) at either 500 or 1000 Hz on an inverted microscope (Nikon TE300 Eclipse) with a Mikrotron high-speed camera (MC 1362) and frame grabber (Inspecta 5, 1 Gb memory) at room temperature (22 $\pm$~$1$~$^{\circ}$C). The images correspond to an area of $\approx 720~\mu\text{m}\times720~\mu\text{m}$, containing around $10^4$ bacteria. Approximately 4000 frames were captured, at a resolution of 512 $\times$ 512 pixels.

To process a video sequence, each frame was divided into square tiles of side length $l$ (typically 5 pixels), and the pixel values in each tile were summed to give a single number. This process was repeated for every frame in the video sequence, yielding intensity as a function of time for each tile. The power spectrum of this data was calculated for each tile separately, before averaging over all tiles to give smoothed data for the whole video sequence. The power spectrum is then normalised by the frequency squared to remove any contribution from Brownian motion due the non-motile cells, inherently present in the bacterial suspensions. An example is shown in Fig.~S\ref{DFM}. We identify the first peak as the body rotational frequency $\Omega/2\pi$ in line with previous studies \cite{Lowe1987}.

\section{Viscosity measurements}

The viscosity of PVP~360kD measured using conventional rheometry and DWS micro-rheology (see Materials \& Methods) at different concentrations are shown in Fig.~S8. There is reasonable overlap between the two methodologies at intermediate shear rates.


\newpage

\newpage

\newpage
\begin{table*}[h]
	\caption{Intrinsic viscosity, [$\eta$], and Huggins coefficient, $k_H$, obtained by fitting simultaneously (global fitting) the viscosity data using both Huggins and Kraemer equations.}
	\begin{center}
	\begin{tabular}{c c c c c}
	\hline
	\\
	Table S1              			& $[\eta]$ (dL/g) 		& $k_H$			& $c^*$(g/dL or wt$\%$) 	\\
	\\
	\hline
	\\
	PVP360k			& 1.84$\pm$0.04			& 0.38$\pm$0.02		& 0.55$\pm$0.01\\
	\\
	PVP160k			& 0.72$\pm$0.01			& 0.38$\pm$0.01		& 1.40$\pm$0.02\\
	\\
	PVP40k			& 0.263$\pm$0.003		& 0.38$\pm$0.02		& 3.8$\pm$0.1\\
	\\
	PVP10k			& 0.105$\pm$0.006		& 0.42$\pm$0.08		& 9.5$\pm$0.5\\
	\\
	\hline
	\\
	SD \cite{Schneider74}	& 1.05$\pm$0.02		& 0.38$\pm$0.02		& 0.95$\pm$0.02\\
	\\
	\hline
	\end{tabular}
	\label{table:rheodata}
	\end{center}
\end{table*}

\newpage
\begin{table*}[h] 
	\caption{Parameters obtained from SLS and DLS for PVP360k in water or in motility buffer (MB).}
	\begin{center}
	\begin{tabular}{c c c c c}
	\hline
	\\
	Table S2              				& $M_w$ (g/mol) 		& $A_2$ (mol.l/g$^2)$			& $R_g$ (nm)		& $R_h$ (nm) 	\\
	\\
	\hline
	\\
	PVP360k	in water		& 840$\times 10^3$		& 3.0$\times10^{-7}$			& 56				& 30\\
	\\
	PVP360k	in MB		& 1500$\times 10^3$	& 2.6$\times10^{-7}$			& 79				& 37\\
	\\
	\hline
	\end{tabular}
	\label{table:zimmdata}
	\end{center}
\end{table*}

\newpage

\begin{figure*}[h]
	\begin{center}
	\includegraphics[width=3.25in]{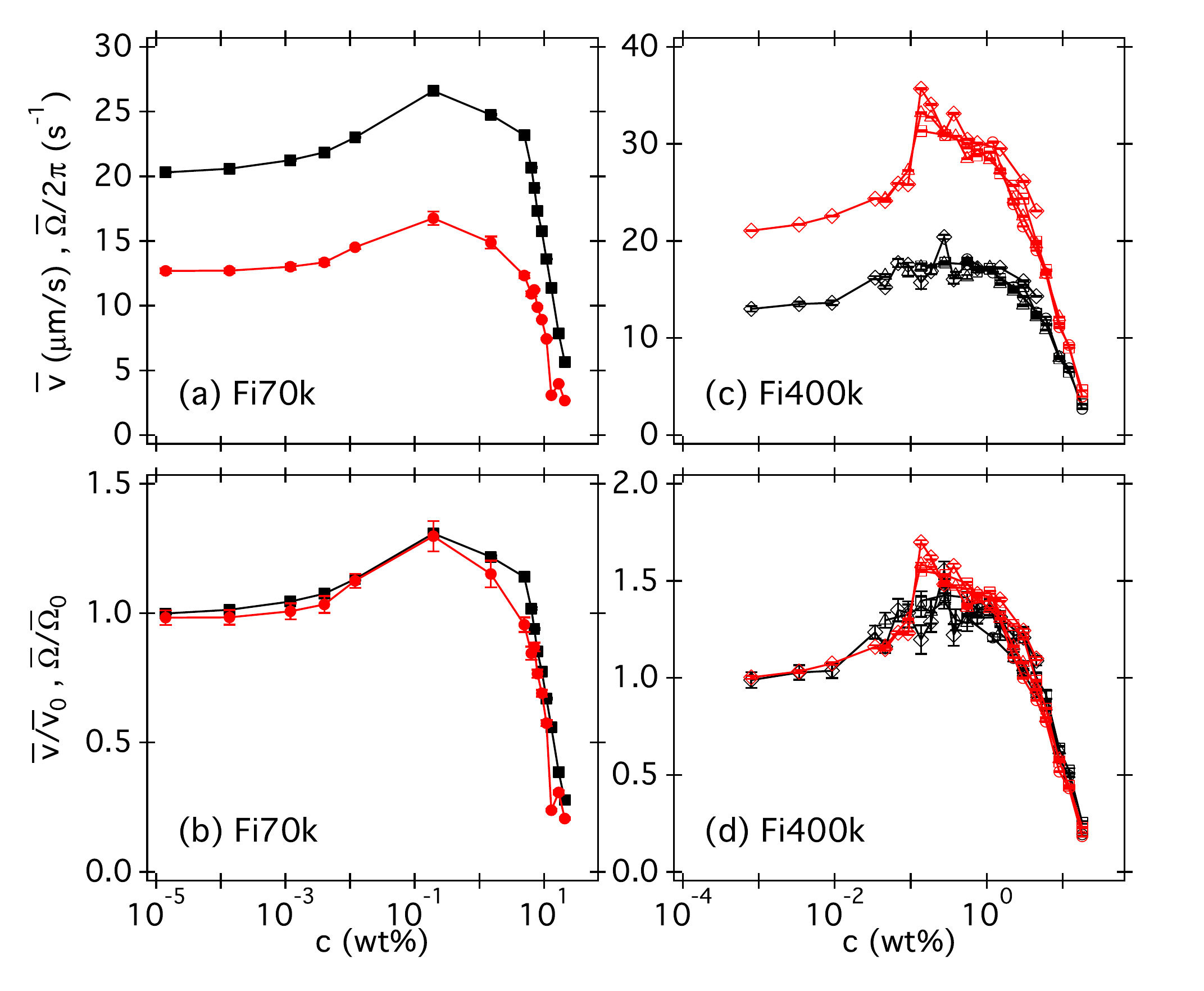}
	\caption{Speed $\bar{v}$ (red symbols) and body rotation frequency $\bar{\Omega}$ (black symbols) given in absolute (top) and  normalised values (bottom) as functions of concentration for native Ficoll of two molecular weights: (a-b) $M$=70k (one dataset) and (c-d) $M$=400k (four datasets). Lines are guides to the eye.}
	\label{dirtyFicoll}
	\end{center}
\end{figure*}

\begin{figure*}[h]
	\begin{center}
	\includegraphics[width=3.5in]{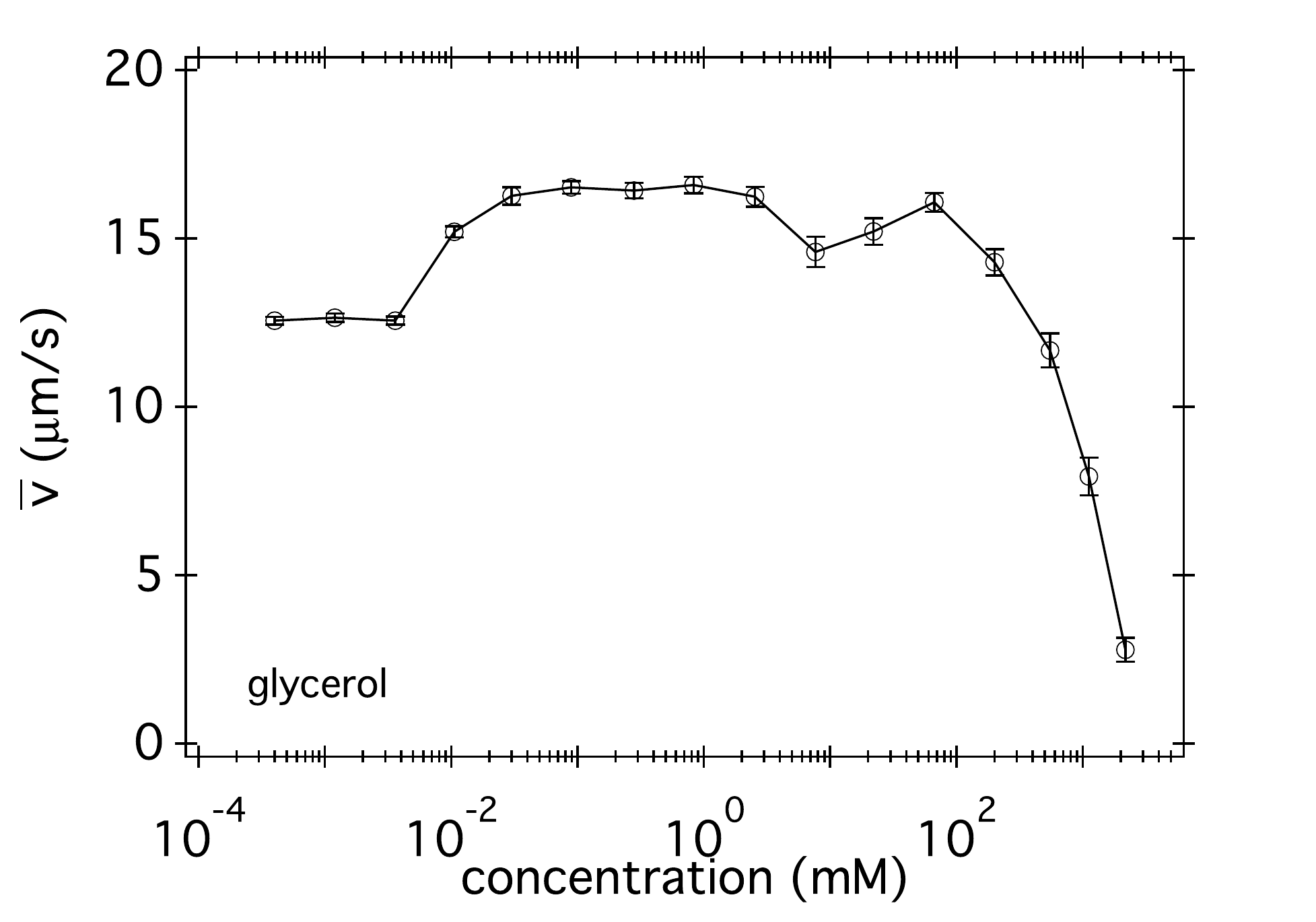}
	\caption{Swimming {\it E. coli} in glycerol. Speed $\bar{v}$ as a function of glycerol concentration.}
	\label{smallmol}
	\end{center}
\end{figure*}

\begin{figure*}[h]
	\begin{center}
	\includegraphics[width=3.25in]{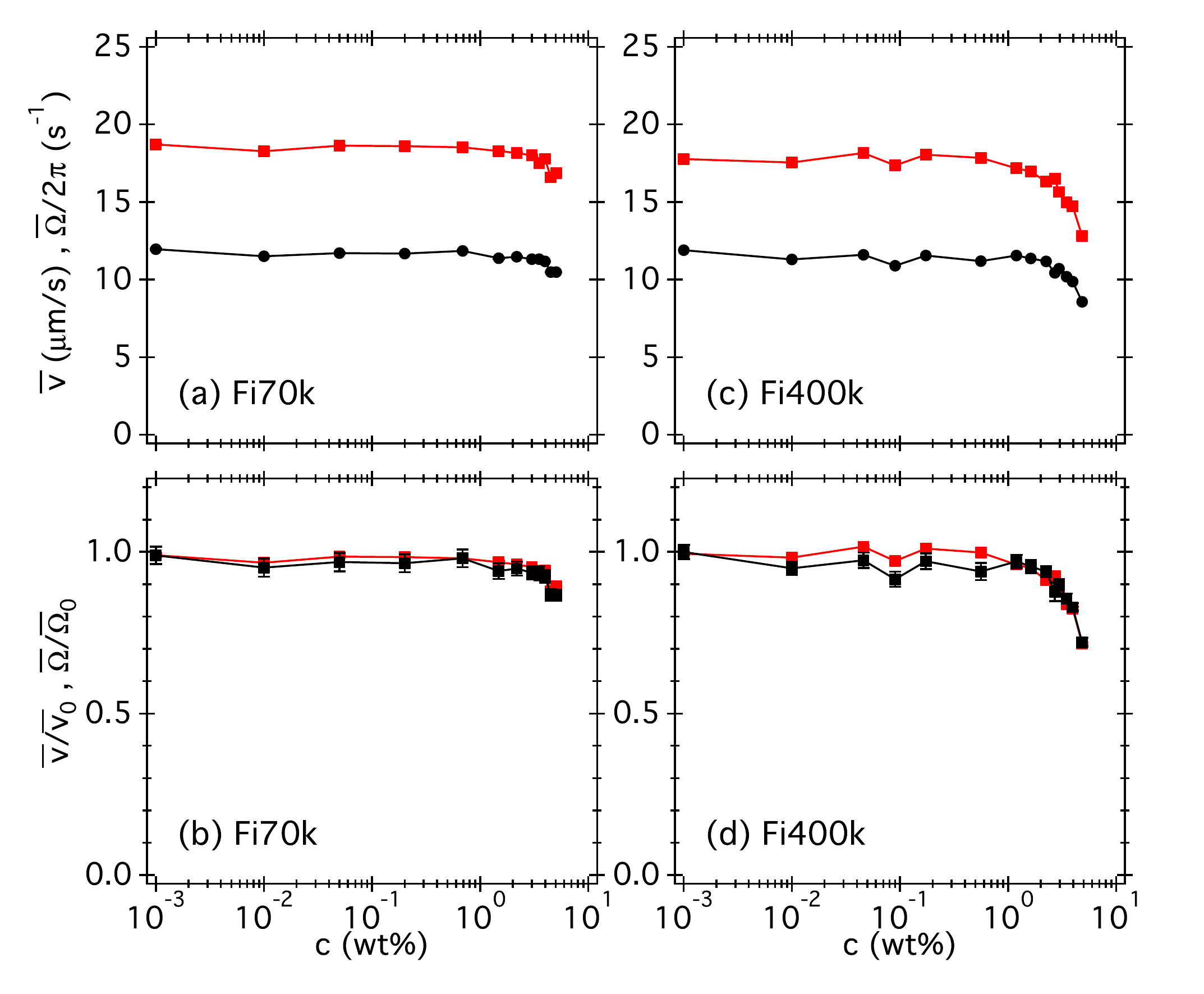}
	\caption{Speed $\bar{v}$ (red circles) and body rotation frequency $\bar{\Omega}$ (black squares) given in absolute (top) and  normalised values (bottom) as functions of concentration for dialysed Ficoll of two molecular weights: (a-b) $M$=70k and (c-d) $M$=400k. Lines are guides to the eye.}
	\label{cleanFicoll}
	\end{center}
\end{figure*}

\begin{figure*}[h]
	\begin{center}
	\includegraphics[width=3.25in]{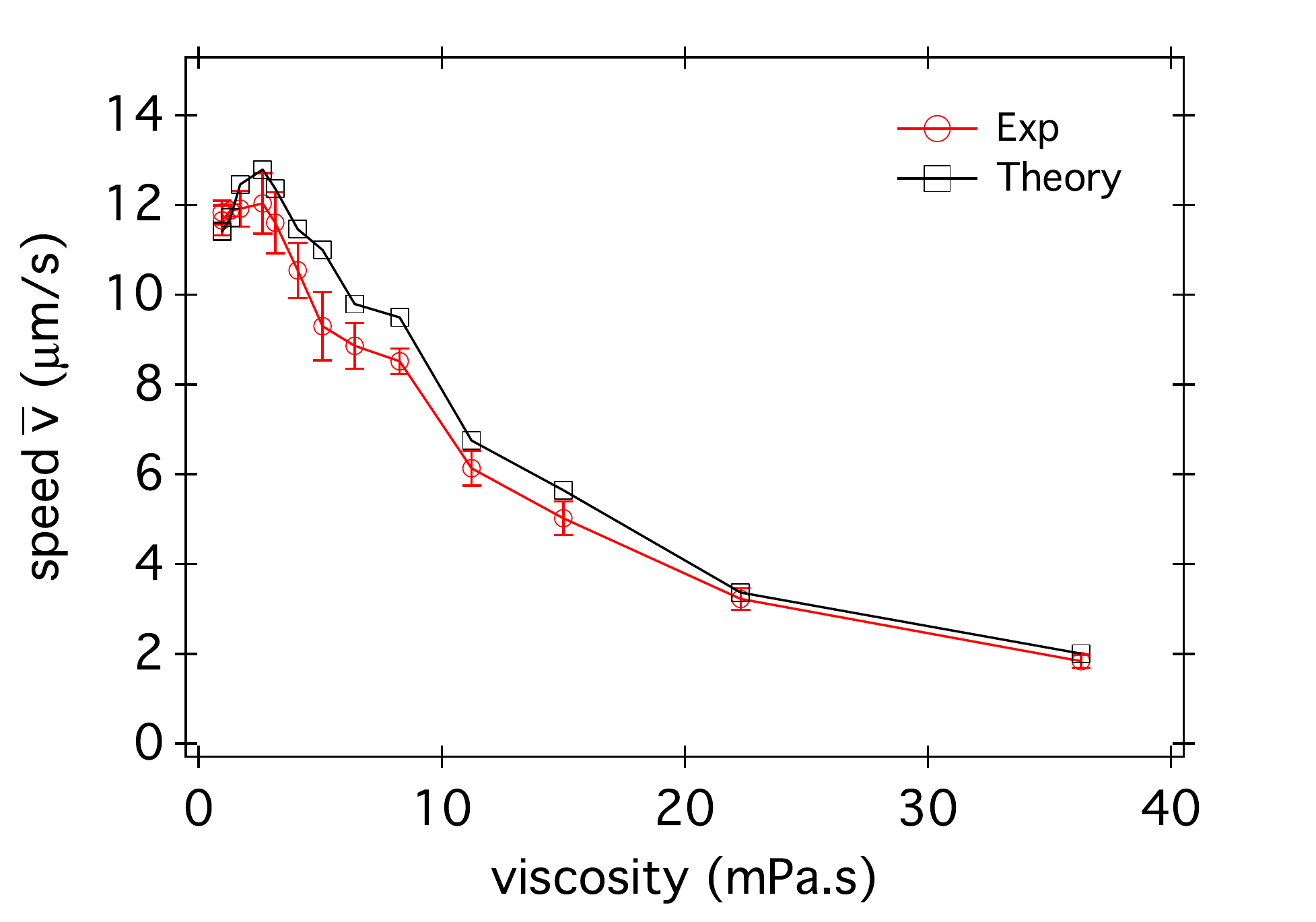}
	\caption{Speed $\bar{v}$ versus viscosity from experiments (black dots)  and our theory (red diamonds) as discussed in the SI text.}
	\label{v_vs_eta}
	\end{center}
\end{figure*}

\begin{figure*}[h]
	\begin{center}
	\includegraphics[width=3.25in]{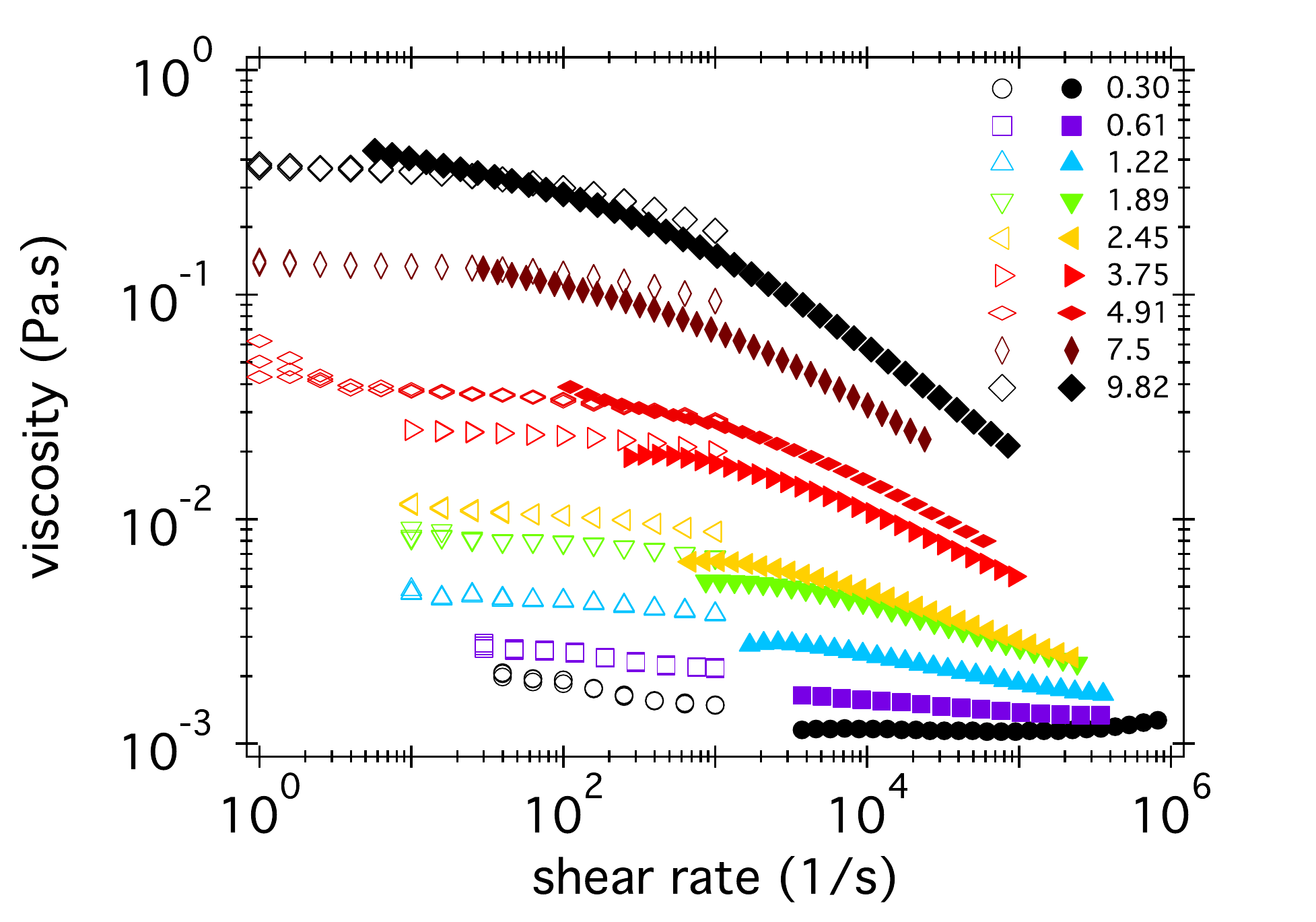}
	\caption{Viscosity of PVP360k as function of shear rate obtained from bulk rheology (open symbols) and DWS micro-rheology (filled symbols) for several polymer concentrations (see legend in weight percent).}
	\label{rheology}
	\end{center}
\end{figure*}

\begin{figure*}[h]
	\begin{center}
	\includegraphics[width=3.25in]{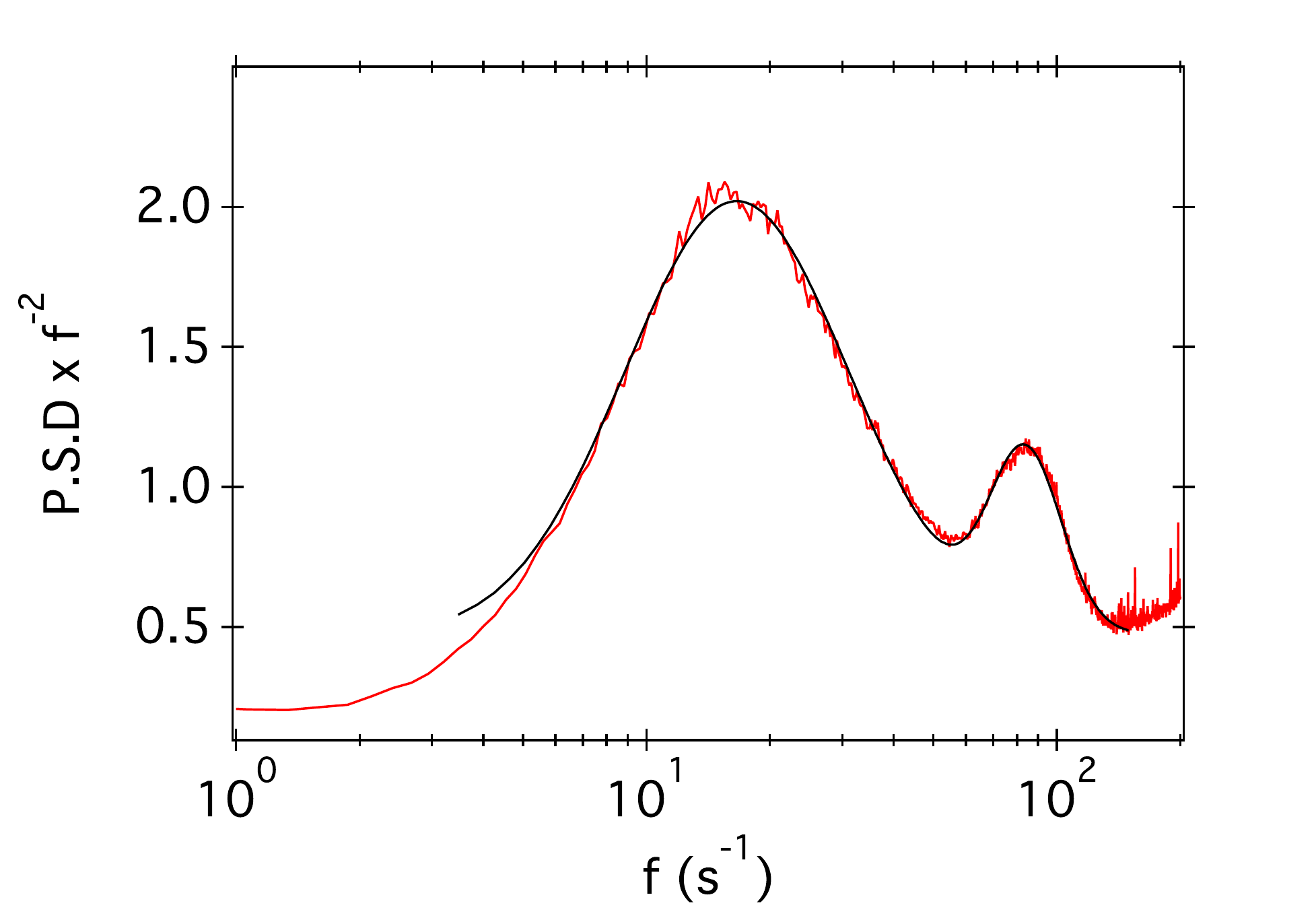}
	\caption{Typical example of the power spectrum of the flickering dark-field image of individual cells averaged over $\approx 10^4$~cells based on a $\approx10$~s dark-field movie (see Materials and Methods). The power spectrum is normalised by the frequency square to remove contribution from Brownian motion due to the inherent presence of non-motile cells in the suspension. The black line corresponds to a two-peak fit using Lognormal distribution.}
	\label{DFM}
	\end{center}
\end{figure*}

\end{article}
\end{document}